\documentclass[a4paper,10pt]{article}
\usepackage[pdftex]{graphicx,hyperref}
\usepackage{amsmath,amsfonts,amssymb}
\usepackage{fullpage}
\usepackage{authblk}
\usepackage{setspace}
\usepackage{subcaption}
\usepackage[explicit]{titlesec}
\usepackage{sidecap}
\usepackage{pbox}
\usepackage{longtable}

\newcommand{\beginsupplement}{%
        \setcounter{table}{0}
        \renewcommand{\thetable}{S\arabic{table}}%
        \setcounter{figure}{0}
        \renewcommand{\thefigure}{S\arabic{figure}}%
     }

\title{\bf{Multi-scale spatio-temporal analysis of human mobility}}

\author[a]{Laura Alessandretti}
\author[b]{Piotr Sapiezynski} 
\author[b,c]{Sune Lehmann}
\author[a,*]{Andrea Baronchelli}

\affil[a]{{\small City,  University of London, London EC1V 0HB, United Kingdom}}
\affil[b]{{\small Technical University of Denmark, DK-2800 Kgs. Lyngby, Denmark}}
\affil[c]{{\small Niels Bohr Institute, University of Copenhagen, DK-2100 K\o benhavn \O , Denmark}

{\small Corresponding authors:  $^{*}$a.baronchelli.work@gmail.com}}

\date{}

\begin{document}
\maketitle

\begin{abstract}
The recent availability of digital traces generated by phone calls and online logins has significantly increased the scientific understanding of human mobility. Until now, however, limited data resolution and coverage have hindered a coherent description of human displacements across different spatial and temporal scales. Here, we characterise mobility behaviour across several orders of magnitude by analysing  $\sim850$ individuals' digital traces sampled every $\sim16$~seconds for 25~months with $\sim10$~meters spatial resolution. We show that the distributions of distances and waiting times between consecutive locations are best described by log-normal and gamma distributions, respectively, and that natural time-scales emerge from the regularity of human mobility. We point out that log-normal distributions also characterise the patterns of discovery of new places, implying that they are not a simple consequence of the routine of modern life. 
\end{abstract}

\section*{Introduction}

\begin{figure}
\centering
\includegraphics[width=\linewidth]{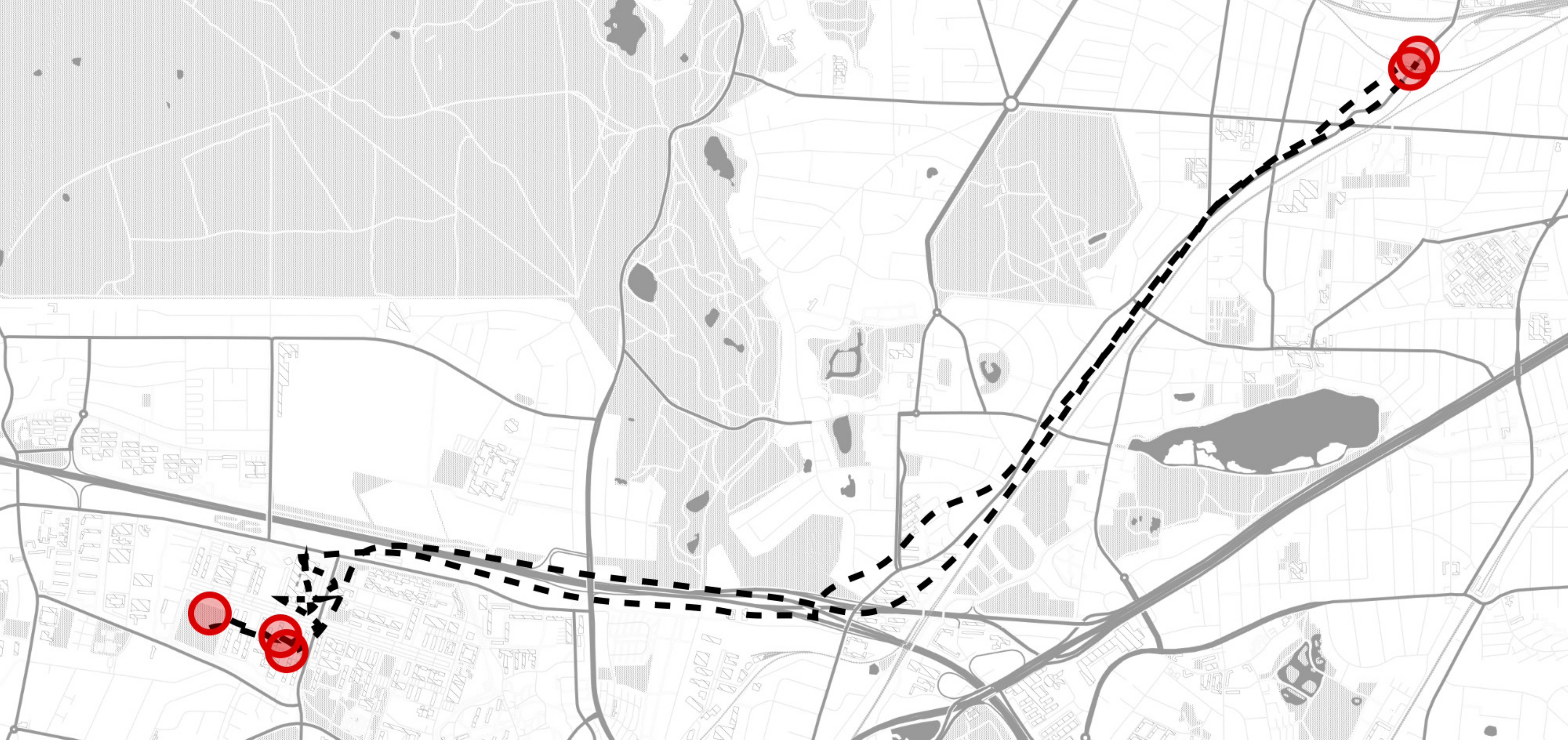}
\caption{\textbf{Example of an individual trajectory}. An individual trajectory is composed of pauses (red dots) and displacements (dashed black line). The trajectory shows the positions of one individual across 26 hours. Location is estimated from individual's WiFi scans as detailed in the text and the data is sampled in 1 min bins. Red dots correspond to locations where the individual spent more than 10 consecutive minutes. The coordinates of these locations have been slightly altered to protect the subject privacy. The map was generated with the Matplotlib Basemap toolkit for Python (\url{https://pypi.python.org/pypi/basemap}). Map data © OpenStreetMap contributors (License: \url{http://http://www.openstreetmap.org/copyright}). Map tiles by Stamen Design, under CC BY 3.0.}
\label{trajectory}
\end{figure}

Characterising the statistical properties of individual trajectories is necessary to understand the underlying dynamics of human mobility and design reliable predictive models. 
A trajectory consists of \emph{displacements} between locations and \emph{pauses} at locations, where the individual stops and spends time (Fig~\ref{trajectory}). Thus, the distribution of waiting times (or pause durations), $\Delta t$, between movements and the distribution of distances, $\Delta r$, travelled between pauses are often used to quantitatively assess the dynamics of human mobility. For example, specific probability distributions of distances and waiting times characterise different types of diffusion processes. Thanks to the recent availability of data used as proxy for human trajectories including mobile phone call records (CDR), location based social networks (LBSN) data, and GPS trajectories of vehicles, the characteristic distributions of distances and waiting times between consecutive locations have been widely investigated. There is no agreement, however, on which distribution best describes these empirical datasets. 

Pioneer studies, based on CDR~\cite{song2010modelling, gonzalez2008understanding} and banknote records~\cite{brockmann2006scaling}, found that the distribution of displacement $\Delta r$ is well approximated by a power-law, $P(\Delta r) \sim \Delta r ^{-\beta}$, (or `L\'evy distribution'\cite{baronchelli2013levy}, as typically $1<\beta < 3$), and that an exponential cut-off in the distribution may control boundary effects \cite{gonzalez2008understanding}. These findings were confirmed by studies based on GPS trajectories of individuals~\cite{wang2014correlations, zhao2015explaining,rhee2011levy} and vehicles~\cite{jiang2009characterizing,liu2012understanding}, as well as online social networks data~\cite{beiro2016predicting,cheng2011exploring,hawelka2014geo}. It has been noted, however, that power-law behaviour may fail to describe intra-urban displacements~\cite{noulas2012tale}. Other analyses, based on online social network data~\cite{wu2014intra,liu2014uncovering, jurdak2015understanding} and GPS trajectories~\cite{liu2015crossover,liang2012scaling,gong2016inferring,zhao2015automatic} showed that the distribution of displacements is well fitted by an exponential curve, $P(\Delta r) \sim e^{-\lambda \Delta r}$, in particular at short distances. Finally, analyses based on GPS on Taxis~\cite{wang2015comparative,tang2015uncovering} suggested that displacements may also obey log-normal distributions, $P(\Delta r) \sim (1/\Delta r) * e^{-(\log \Delta r - \mu)^2/2 \sigma^2} $. In Ref.~\cite{zhao2015explaining}, the authors found that this is the case also for single-transportation trips.

Fewer studies have explored the distribution of waiting times between displacements, $\Delta t$, as trajectory sampling is often uneven (e.g., in CDR data location is recorded only when the phone user makes a call or texts, and LBSN data include the positions of individuals who actively ``check-in" at specific places). Analyses based on evenly sampled trajectories from mobile phone call records~\cite{song2010modelling,schneider2013unravelling}, and individuals GPS trajectories~\cite{wang2014correlations,rhee2011levy} found that the distribution of waiting times can be also approximated by a power-law.
A recent study based on GPS trajectories of vehicles, however, suggests that for waiting times larger than $4$ hours, this distribution is best approximated by a log-normal function~\cite{gallotti2016stochastic}. Several studies have highlighted the presence of natural temporal scales in individual routines: distributions of waiting times display peaks in that corresponds to the typical times spent home on a typical day ($\sim14$~hours) and at work ($\sim3-4$~hours for a part-time job and $\sim8-9$~hours for a full-time job)\cite{schneider2013unravelling,hasan2013spatiotemporal,bazzani2010statistical}.

\begin{figure}
\centering
\includegraphics[width=\linewidth]{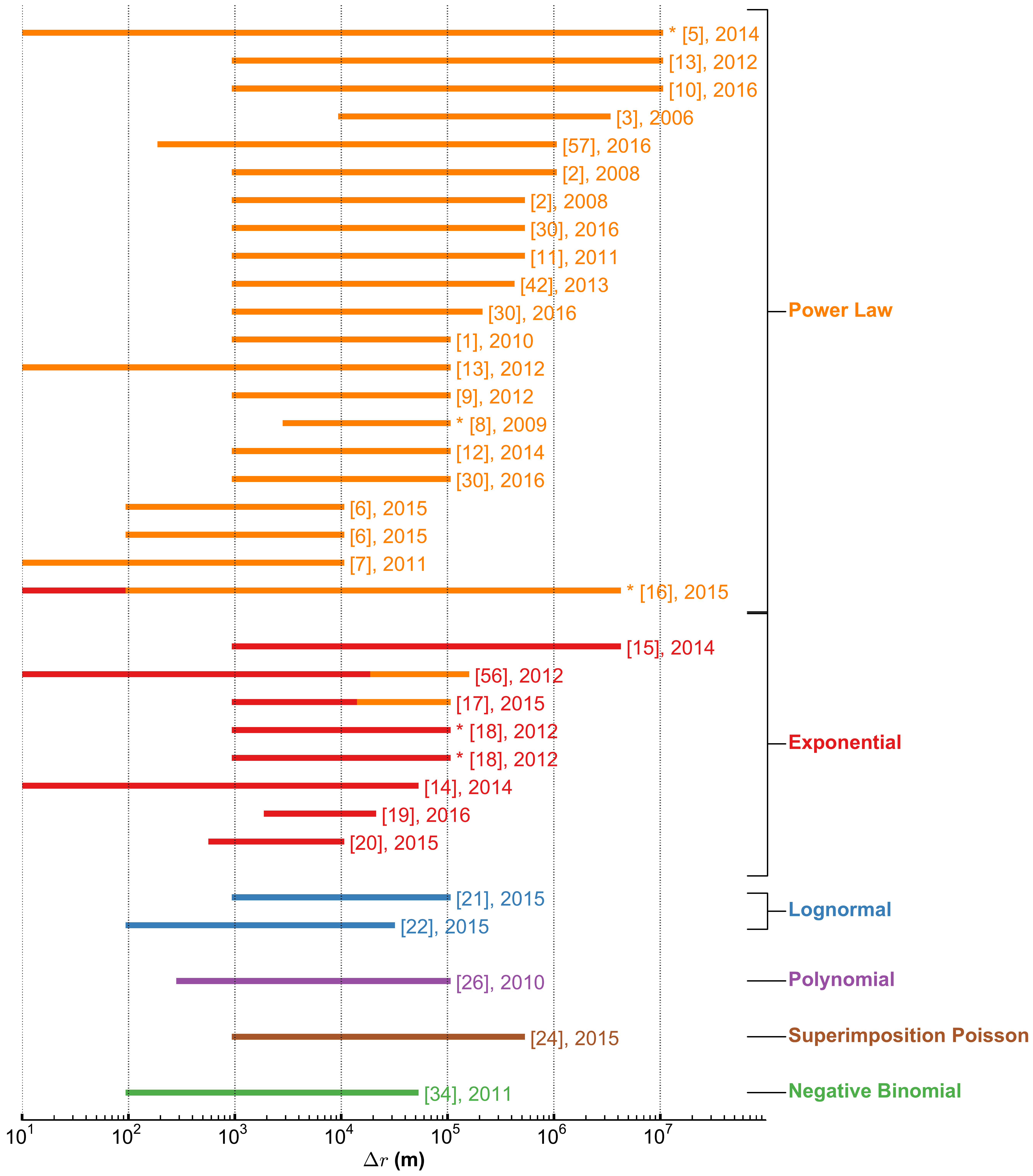}
\caption{ \textbf{The distribution of displacements $P(\Delta r)$: heterogeneity of results found in the literature.} Each horizontal line corresponds to a different dataset. Lines extend from the minimum $\Delta r$ (\emph{i.e.} the spatial resolution of the data or the minimum value considered for the fit of the distribution), to the maximal length of displacement considered (both in meters). Colours correspond to the model fitting $P(\Delta r)$ according to the study reported at the end of each line. If the distribution is not unique, but varies for different ranges of $\Delta r$, the line is divided in segments. Lines are marked with `*' if the corresponding data is modelled as a sequence of two distributions of the same type with different parameters, for different ranges $\Delta r$. Refs \cite{gonzalez2008understanding,liang2012scaling,deville2016scaling,zhao2015explaining} analyse more than one dataset. In \cite{noulas2012tale} the authors analyse the same dataset for different ranges $\Delta r$. A more detailed table is presented in section ``Related Work".}
\label{Fig_ranges}
\end{figure}

Fig~\ref{Fig_ranges} and Table \ref{theTable} compare distributions obtained using different data sources. The spectrum of results reflects the heterogeneity of the considered datasets (see Fig~\ref{Fig_ranges}). It is known in fact that data spatio-temporal resolution and coverage has an important influence on the results of the analyses performed \cite{paul2016scaling,decuyper2016clean,kivela2015estimating}.

First, the datasets considered have different \emph{spatial resolution and coverage}, and few studies have so far considered the whole range of displacements occurring between $\sim10$ and $10^7~\mathrm{m}$ ($10000~\mathrm{km}$) (Fig~\ref{Fig_ranges}). Our analysis suggests that constraining the analysis to a specific distance range may result in different interpretations of the distributions.
Another difference concerns the \emph{temporal sampling} in the datasets analysed so far. Uneven sampling typical of CDR and LBSN data (i) does not allow to distinguish phases of \emph{displacement} and \emph{pause}, since individuals could be active also while transiting between locations, and (ii) may fail to capture patterns other than regular ones~\cite{ccolak2015analyzing,ranjan2012call}, because individuals' voice-call/SMS/data activity may be higher in certain preferred locations.
Finally, studies focusing on displacements effectuated using one or several \emph{specific transportation modality} (private car~\cite{gallotti2016stochastic,gallotti2015understanding}, taxi ~\cite{zhao2015automatic}, public transportation~\cite{roth2011structure}, or walk~\cite{rhee2011levy}) capture only a specific aspect of human mobility behaviour.

In this paper, we analyse mobility patterns of $\sim 850$ individuals involved in the Copenhagen Network Study experiment for over $2$~years~\cite{stopczynski2014measuring}. Individual trajectories are determined combining GPS and Wi-Fi scans data resulting in a spatial resolution of $\sim10~\mathrm{m}$, and even sampling every $\sim16~\mathrm{s}$. Trajectories span more than $\sim10^7~\mathrm{m}$. Previous studies with comparable spatial coverage (Fig~\ref{Fig_ranges}) relied on single-transportation modality data~\cite{jiang2009characterizing}, unevenly sampled data~\cite{jurdak2015understanding}, or small samples (32 individuals in Ref.~\cite{wang2014correlations}). To our knowledge, the Copenhagen Network Study data has the best combination of spatio-temporal resolution and sample size among the datasets analysed in the literature to date (see Methods).

\section*{Results}

We consider an individual to be \emph{pausing} when he/she spends at least 10 consecutive minutes in the same location, and \emph{moving} in the complementary case. In the following, we refer to \emph{locations} as places where individuals pause. The distribution of displacements is robust with respect to variations of the pausing parameter (see Supplementary Information for the results obtained with 15 and 20 minutes pausing). 

We start by considering the three distributions most frequently reported in the literature (Table~\ref{theTable}), namely

\begin{itemize}
\item \emph{The log-normal distribution} of a random variable $x$, with parameters $\sigma$ and $\mu$, defined for $\sigma > 0$ and $ x > 0 $, with probability density function: 
\begin{equation}\label{eq:lognorm}
P(x) = \dfrac{1}{\sqrt{2 \pi \sigma^2} x } e^{-\dfrac{1}{2} \dfrac{(\log x - \mu)^2}{\sigma^2}}
\end{equation}
\item \emph{The Pareto distribution} (\emph{i.e.} power-law) of a random variable $x$, with parameter $\beta$, defined for $x \geq 1 $, and $\beta > 1$, with probability density function: 
\begin{equation}\label{eq:pareto}
P(x) = (\beta-1) \left( x \right)^{-\beta}
\end{equation}

\item \emph{The exponential distribution} of a random variable $x$, with parameter $\lambda$, where $ x \geq 0 $, and $\lambda > 0$, with probability density function: 
\begin{equation}\label{eq:expon}
P(x) = \lambda e^{-\lambda x}
\end{equation}
\end{itemize}
\noindent In equation (\ref{eq:pareto}) the probability density can be shifted by $x_0$ and/or scaled by $s$, as $P(x)$ is identically equivalent to $P(y)/s$, with $y = \dfrac{(x-x_0)}{s}$. In equations (\ref{eq:lognorm}), and (\ref{eq:expon}), $P(x)$ is identically equivalent to $P(y)$, with $y = (x-x_0)$. In this work, the shift ($x_0$) and scale ($s$) parameters are considered as additional parameters to take into account the data resolution. With few exceptions, the results presented below hold also imposing no shift, $x_0 = 0$ (see Supplementary Information). Note also that 
Pareto distributions with exponential cut-off (or truncated Pareto) are considered below
(see also table \ref{theTable}).

\subsection*{Distribution of displacements}

\begin{figure}
\centering
\includegraphics[width=\linewidth]{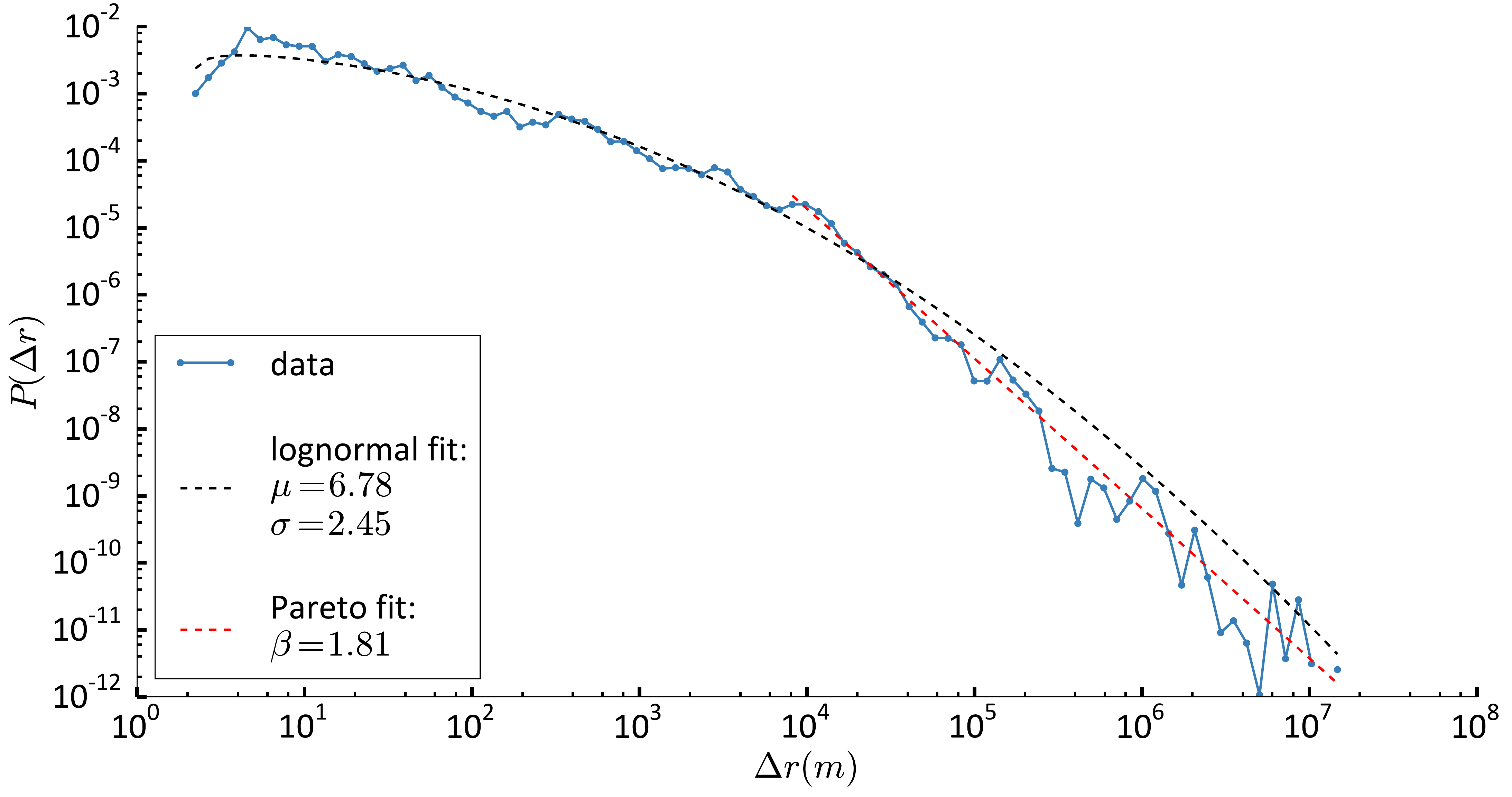}
\caption{\textbf{Distribution of displacements}. Blue dotted line: data. Black dashed line: log-normal fit with characteristic parameter $\mu$ and $\sigma$. Red dashed line: Pareto fit with characteristic parameter $\beta$ for $\Delta r>7420~\mathrm{m}$.}
\label{distance_between_locations}
\vspace{0.8cm}
\end{figure}

We start our analysis by investigating the distribution of displacements between consecutive stop-locations $P(\Delta r)$. First, we consider the overall distribution of the displacements $\Delta r$ using all available data (851 individuals over 25 months). We find that $P(\Delta r)$ is best described by a log-normal distribution (equation \ref{eq:lognorm}) with parameters $\mu=6.78 \pm 0.07$ and $\sigma=2.45 \pm 0.04 $, which maximises Akaike Information Criterion (see Methods) --- among the three models considered --- with Akaike weight $\sim 1$ (Fig~\ref{distance_between_locations}, see also SI). 

Second, we investigate if this results holds also for sub-samples of the entire dataset. We bootstrap data $1000$ times for samples of $200$ and $100$ individuals, and we verify that the best distribution is log-normal for all samples, and the average parameters inferred through the bootstrapping procedure are consistent with the parameters found for the entire dataset (see the Supplementary Information). In fact, the errors on the value of the parameters reported above are computed by bootstrapping data for samples of $100$ randomly selected individuals. This analysis ensures homogeneity within the population considered, and takes into account also that often smaller sample sizes were analysed in previous literature.

\begin{figure}
\centering
\includegraphics[width=\linewidth]{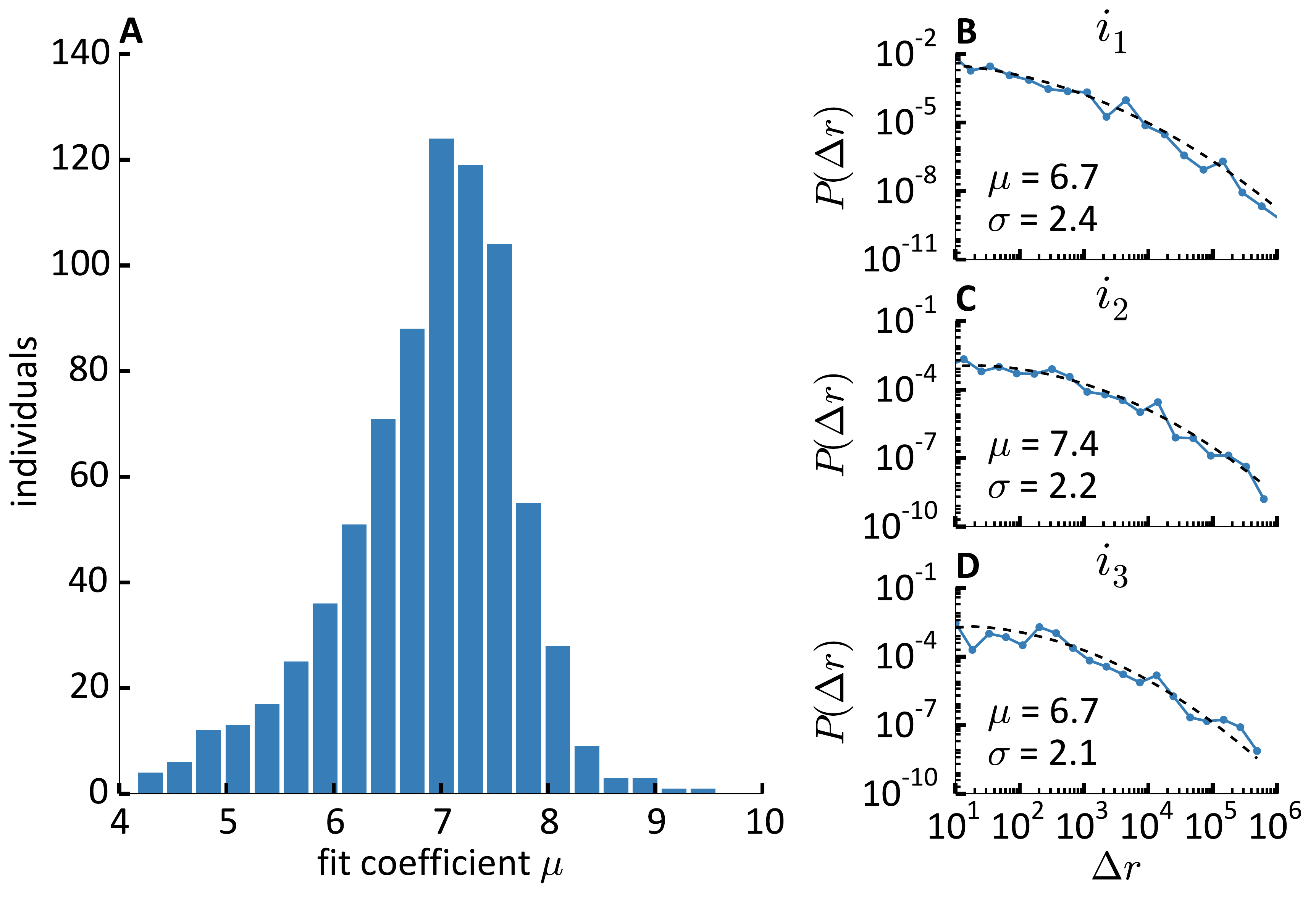}
\caption{\textbf{Distribution of individual displacements}. \textbf{A}) Frequency histogram of 96.2\% of individuals for which the individual distribution of displacement is log-normal, according to the value of the log-normal fit coefficient $\mu$. \textbf{B}-\textbf{C}-\textbf{D}) Examples of the distribution of displacements $P(\Delta r)$ of three individuals $i_1$ (B), $i_2$ (C), $i_3$ (D) (dotted line), with the corresponding log-normal fit (dashed line). The value of the fit coefficients $\mu$ and $\sigma$ are reported in each subfigure. }
\label{Fig_individuals}
\vspace{0.8cm}
\end{figure}

Third, we zoom in to the individual level. We find that the individual distribution of displacements is best described by a log-normal function for $96.2\%$ of individuals. The best distribution is the Pareto distribution for $1.4\%$, and exponential for the remaining $2.4\%$. However, the number of data points per individual tend to be significantly lower in group of individuals exhibiting Pareto or exponential distributions, so that one should be cautious in interpreting the observed deviations from a log-normal distribution. Fig~\ref{Fig_individuals} reports the histogram of the individual $\mu$ parameters for the $96.2\%$ of the population that is best described by a log-normal distribution, along with three examples of individual distributions. 

Finally, we look at large $\Delta r$ in order to compare our results with precedent studies relying on data with larger spatial resolution. We find that limiting the analysis to large values of $\Delta r$ results in the selection of a Pareto distribution (equation \ref{eq:pareto}). We identify the threshold $\Delta r*=7420~\mathrm{m}$ as the minimal resolution for which the best fit in $\Delta r* <\Delta r<10^7~\mathrm{m}$ is Pareto with coefficient $\beta = 1.81 \pm 0.03$ and not log-normal. By bootstrapping $1000$ times over samples of $100$ individuals we find that $\hat{\Delta r*}=7488.3 \pm 328.2~\mathrm{m}$. Thus, power-law distributions describe mobility behaviour only for large enough distances, while mobility patterns including distances smaller than $7420~\mathrm{m}$ are better described by log-normal distributions.

\subsection*{Distribution of waiting times}

\begin{figure}
\centering
\includegraphics[width=\linewidth]{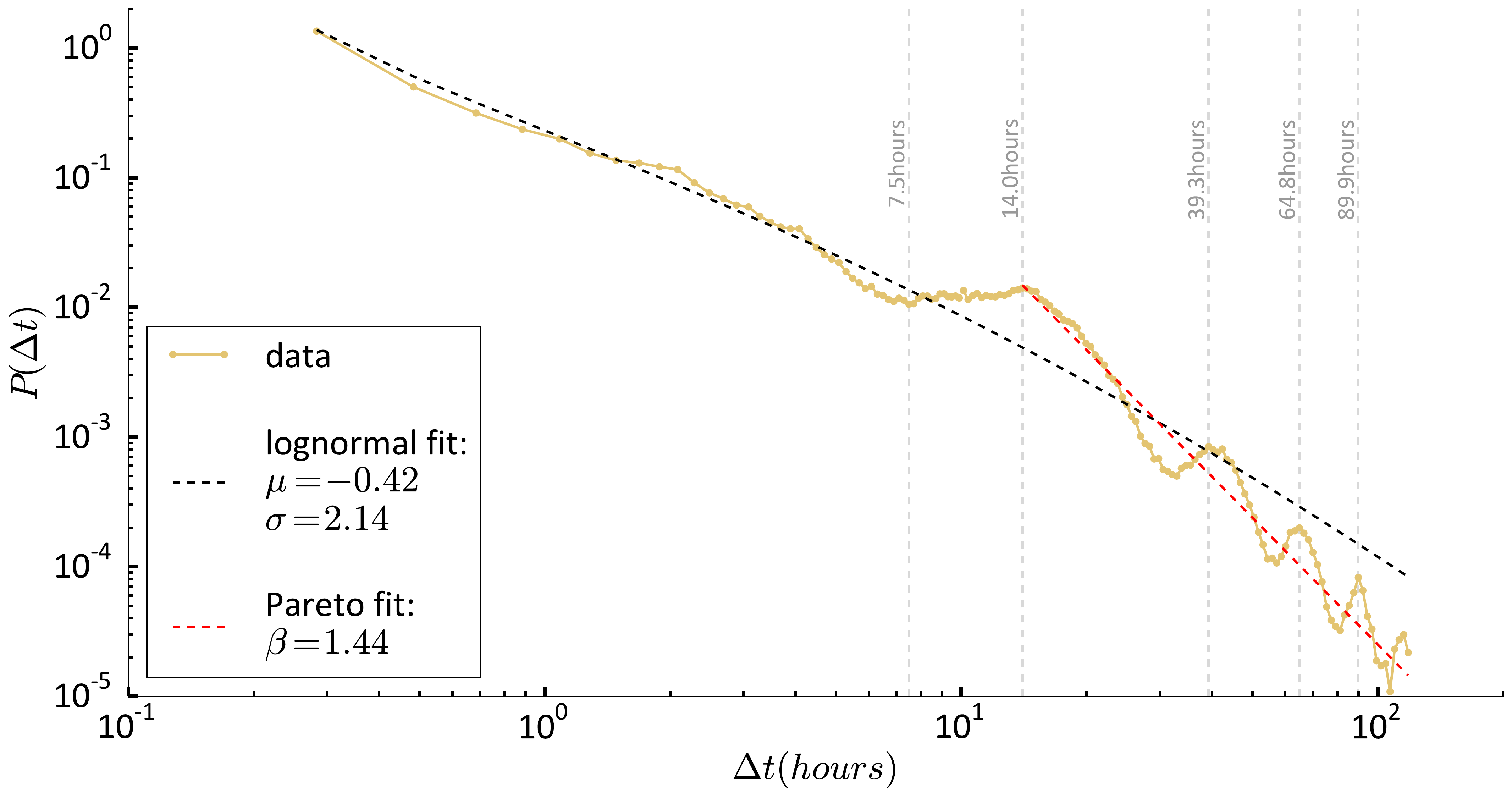}
\caption{\textbf{Distribution of waiting times between displacements.} Yellow dotted line: data. Black dashed line: Log-normal fit with characteristic parameter $\mu$ and $\sigma$. Red dashed line: Pareto fit with characteristic parameter $\beta$ for $\Delta t>13~\mathrm{h}$.}
\label{waiting_times between locations}
\vspace{0.8cm}
\end{figure}

We now analyse the distribution of waiting times between displacements. The best model describing the distribution of waiting times over all individuals is the log-normal distribution (equation \ref{eq:lognorm}, Fig~\ref{waiting_times between locations}, see also SI), with parameters $\mu=-0.42 \pm 0.04$, $\sigma=2.14 \pm 0.02$. As above, errors are found by bootstrapping over samples of $100$ individuals. Also, by bootstrapping we find that the log-normal distribution is the best descriptor for samples of 200 and 100 randomly selected individuals (see Supplementary Information). As in the case of displacements, we find that restricting the analysis to large values of our observable $\Delta t$, and specifically considering only $\Delta t>\Delta t*=13~\mathrm{h}$, results in the selection of the Pareto distribution (equation \ref{eq:pareto}, see Fig~\ref{waiting_times between locations}), with coefficient $\beta = 1.44 \pm 0.01$. We find by averaging over 100 samples of 200 individuals that $\hat{\Delta t*}=13.01 \pm 0.12$. Note that the log-normal distribution is selected as the best model also when the analysis is restricted to $\Delta t< \Delta t*$.

The distribution of waiting times shows also the existence of ``natural time-scales" of human mobility. We detect local maxima of the distribution at 14.0, 39.3, 64.8, and 89.9 hours. Hence, 14 hours is the typical amount of time that students in the experiment spent home every day, in agreement with previous analyses on human mobility~\cite{schneider2013unravelling,hasan2013spatiotemporal,bazzani2010statistical}. Other peaks appear for intervals $\Delta t \approx 14+n \cdot 24$, with $n = \{2,3...\}$, suggesting individuals spend several days at home. Notice also that the distribution we consider is limited to $\Delta t < 5$ days, an interval much shorter than the observation time-window (about 2 years), a fact that guarantees the absence of possible spurious effects\cite{kivela2015estimating}. This limit is imposed to  control the cases in which students leave their phones home. The upper bound is arbitrarily set to 5 days; however, we have verified that results are consistent with respect to variations of this choice.

\subsection*{Distribution of displacements between discoveries}

\begin{figure}
\centering
\includegraphics[width=\linewidth]{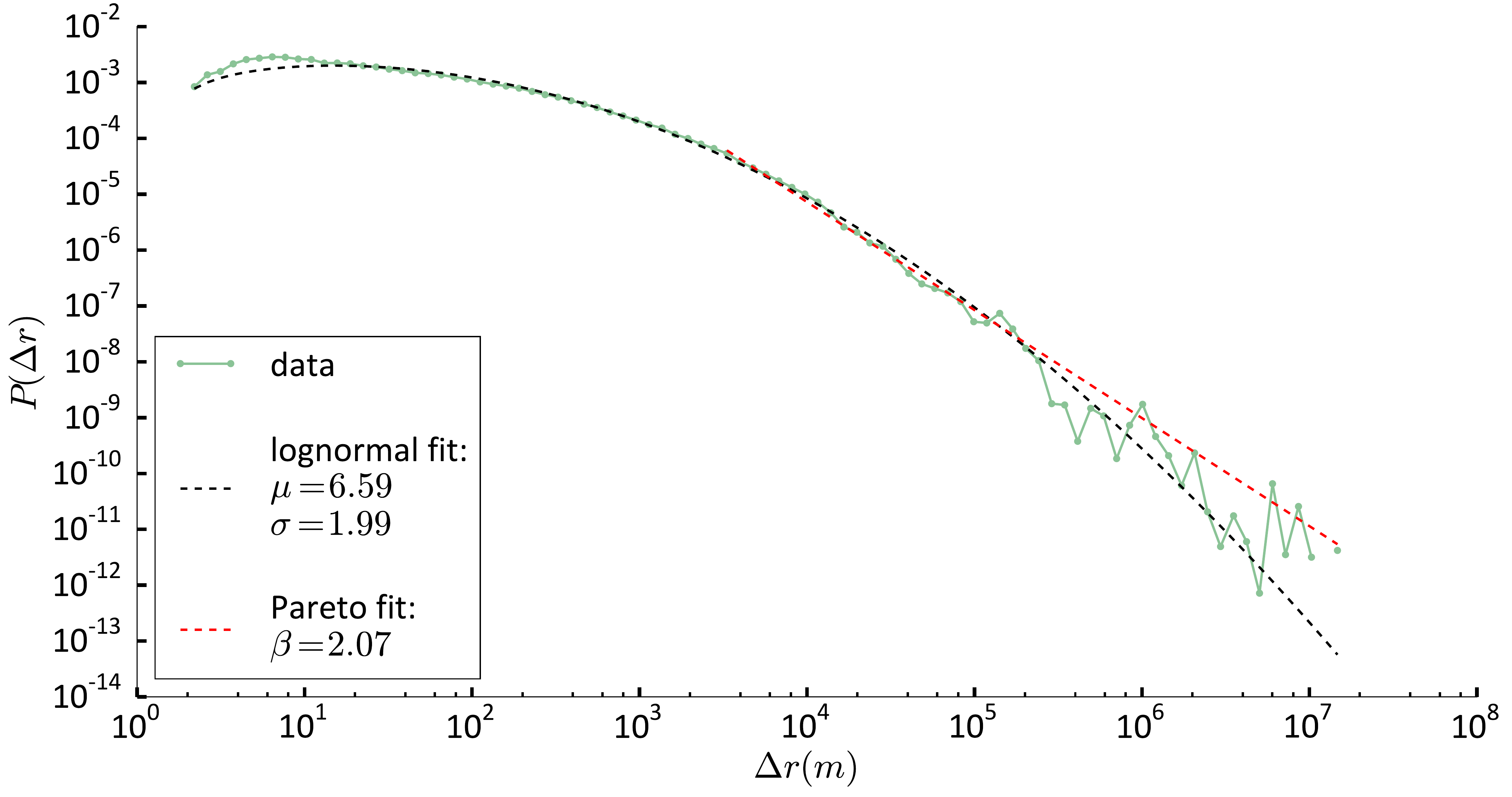}
\caption{\textbf{Distribution of displacements between discoveries}. Green dotted line: data. Black dashed line: Log-normal fit with characteristic parameter $\mu$ and $\sigma$. Red dashed line: Pareto fit with characteristic parameter $\beta$ for $\Delta r>2800~\mathrm{m}$.}
\label{distance_between discoveries}
\vspace{0.8cm}
\end{figure}

Log-normal features also characterise patterns of \textit{exploration}. We consider the temporal sequence of stop-locations that individuals visit for the first time --- in our observational window --- and characterise the distributions of displacements between these `discoveries'. 
We find that the distribution of distances between consecutive discoveries $P(\Delta r)$ is best described as a log-normal distribution with parameters $\mu=6.59 \pm 0.02$, $\sigma=1.99 \pm 0.01$, (Fig~\ref{distance_between discoveries},  see also SI). For $\Delta r>2800~\mathrm{m}$, the best model fitting the distribution of displacements is the Pareto distribution with coefficient $\beta = 2.07 \pm 0.02$. This results are verified by bootstrapping (see Supplementary Information).

\noindent

\subsection*{Correlations between pauses and displacements}

\begin{figure}
\centering
\includegraphics[width=\linewidth]{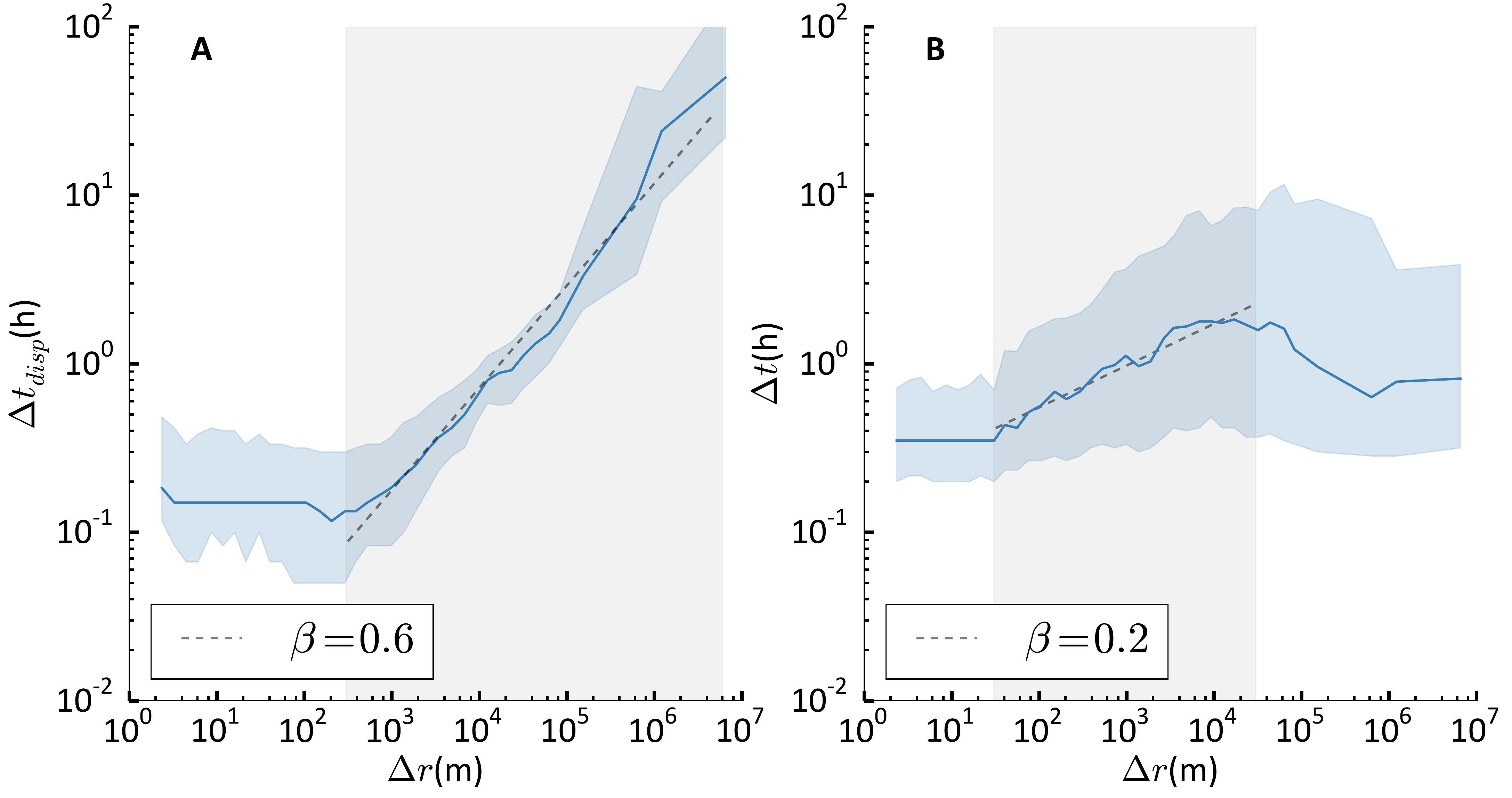}
\caption{\textbf{Correlations between displacements and pauses}. \textbf{A)} The duration $\Delta t_{disp}$ of a displacement vs the distance $\Delta r$ between origin and destination. The blue line is the median value of $\Delta r$ and $\Delta t_{disp}$ computed within log-spaced 2-dimensional bins. The filled blue area corresponds to the  25-75 percentile range. The value of the Pearson correlation coefficient within the shaded grey area indicates a positive correlation, with $p-value<0.01$. The dashed line is a power-law function with coefficient $\beta$, as a guide for the eye. \textbf{B)} The waiting time $\Delta t$ at destination vs the distance $\Delta r$ between origin and destination. The blue line is the median value of $\Delta r$ and $\Delta t$ computed within log-spaced 2-dimensional bins. The filled blue area corresponds to the 25-75 percentile range. The value of the Pearson correlation coefficient within the shaded grey area indicates a positive correlation, with $p-value<0.01$. The dashed line is a power-law function with coefficient $\beta$, as a guide for the eye. }
\label{correlations}
\vspace{0.8cm}
\end{figure}

We further investigate the properties of individual trajectories by analysing the correlations between the distance $\Delta r$ and the duration $\Delta t_{disp}$ characterising a displacement and the time $\Delta t$ spent at destination. Fig~ \ref{correlations}A shows a positive correlation between $\Delta r$ and $\Delta t_{disp}$ for $\Delta r \gtrsim 300 m$ ($p<0.01$). As $\Delta r$ is the distance between the displacement origin and destination, the absence of correlation at short distances could be due to individuals not taking the fastest route.  A positive correlation characterises also the distance $\Delta r$ covered between origin and destination and the waiting time at destination for distances $30 m \lesssim \Delta r \lesssim 10^4 m$ ($p<0.01$). Instead, the correlation is negative for distances larger than $5 \times 10^4m$ (Fig~ \ref{correlations}B). This could suggest that individuals break long trips with short pauses. We have verified that these results hold also when individuals' most important locations (typically including university and home) are removed from the trajectory, implying that these correlations are not dominated by daily commuting.

\subsection*{Further analysis: Selection of the best model among 68 distributions}

In the previous sections we have restricted the analysis of the distributions of displacements and waiting times to the three functional forms that are most frequently found in the literature. We now repeat the selection procedure considering a list of 68 models (see Supplementary Information for the list of distributions) in order to confirm the results described above. 

The distributions of displacements and displacements between discoveries are best described by log-normal distributions also when the choice is extended to 68 models, and tails (respectively for $\Delta r>\Delta r*=7420~\mathrm{m}$ and $\Delta r>\Delta r*= 2800~\mathrm{m} $) are better modelled as generalised Pareto distribution, with form:
\begin{equation}
P(x) = \left( 1 + \xi x \right) ^{-\frac{\xi+1}{\xi}}
\end{equation}
where $\xi$ is the parameters of the model, such that $ x \geq 0$ if $\xi \geq 0$, and $0 \leq x \leq - \frac{1}{\xi}$ if $\xi < 0$. 

The best model selected for the whole distribution of waiting time among the 68 models considered is a gamma distribution, defined for $x \in (0, \infty)$, $k>0$ and $\theta >0$ as:
\[
P(x) = \frac{1}{\Gamma(k) \theta^{k}}{x^{k-1}e^{-\frac{x}{\theta}}}
\]
where $\Gamma(z) = \int_0^\infty x^{z-1} e^{-x} dx$. Although the gamma distribution is the best model for the distribution of waiting times (see SI for the result of the fit), the presence of natural scales could indicate that the whole distribution may be better described as the composition of several models.

\section*{Discussion}

Using high resolution data we have characterised human mobility patterns across a wide range of scales. We have shown that both the distribution of displacements and waiting times between displacements are best described by a log-normal distribution. We found, however, that power-law distributions are selected as the best model when only large spatial or temporal scales are considered, thus explaining (at least partially) the disagreement between previous studies. We also showed that log-normal distributions characterise the distribution of displacements between discoveries, implying that this property is not a simple consequence of the stability of human mobility but a characteristic feature of human behaviour. Finally, we have shown that there exist correlations between displacements' length and the waiting time at destination. 

The heavy tailed nature of human mobility has been attributed to various factors, including differences between individual trajectories~\cite{petrovskii2011variation}, search optimisation~\cite{viswanathan1999optimizing,lomholt2008levy,raposo2009levy,santos2007origin}, the hierarchical organisation of the streets network~\cite{han2011origin} and of the transportation system~\cite{zhao2015explaining, gallotti2016stochastic, yan2013diversity}. On the other hand log-normal distributions can result from multiplicative~\cite{mitzenmacher2004brief} and additive~\cite{mouri2013log} processes and describe the inter-event time of different human activities such as writing emails, commenting/voting on online content \cite{van2011lognormal} and creating friendship relations on online social networks~\cite{blenn2016human}. Instead, the distribution of inter-event time in mobile-phone call communication activity can be described as the composition of power-laws \cite{karsai2011small, jo2012circadian, krings2012effects}, a feature attributed to the existence of characteristic scales in communication activity such as the time needed to answer a call, as well as the existence of circadian, weakly and monthly patterns. We also find clear signatures of circadian patterns, which could indicate that the whole distribution may be better described as the composition of several models. However, in our case the best description for times including $ \Delta t < \Delta t^*$ is  the gamma distribution, which thus is selected both when the whole range of scales is considered and when the analysis is restricted to short times. 

Our results come from the analysis of a sample of $\sim 850$ University students, which of course represent a very specific sample of the whole population. Nevertheless, it is worth noting that many statistical properties of CNS students mobility patterns are consistent with previous results, such as the distribution of the radius of gyration, the Zipf-like behaviour of individual locations frequency-rank plot, and the power-law tail of the distribution of displacements 
($\beta = 1.81 \pm 0.03$ vs. $\beta = 1.75 \pm 0.15$ of \cite{gonzalez2008understanding}). Details are reported in Supplementary Information of \cite{1609.03526}.\\

While identifying the mechanism responsible for the observed mobility patterns is beyond the scope of the present article, we anticipate that a more complete spatio-temporal description of human mobility will help us develop better models of human mobility behaviour~\cite{gallotti2016stochastic, gutierrez2015active}. Our findings can also help the understanding of phenomena such as the spreading of epidemics at different spatial resolutions, since the nature of heterogeneous waiting times between displacements have a major impact on the spreading of diseases~\cite{poletto2013human}. 

\section*{Methods}

\subsection*{Data description and pre-processing}
The Copenhagen Network Study data collection took place between September $2013$ and February $2016$ and involved $851$ students of Technical University of Denmark (DTU) in Copenhagen. Data collection was approved by the Danish Data Protection Agency. All participants provided informed consent by filling an on-line consent form and all methods were performed in accordance with the relevant guidelines and regulations. Individual trajectories were inferred combining WiFi scans data and GPS scans data recorded on smartphones handed out to all participants. An anthropological field study included in the 2013 deployment of the experiment reported that participants did not alter their habits due to participation in the CNS experiment.

The WiFi scans data provides a time-series of wireless network scans performed by participants' mobile devices. Each record \textit{(i, t, SSID, BSSID, RSSI)} indicates:
\begin{itemize}
\item{the participant identifier, \textit{i}}
\item{the timestamp in seconds, $t$}
\item{the name of the wireless network scanned, \textit{SSID}}
\item{the unique identifier of the access point (AP) providing access to the wireless network, \textit{BSSID}}
\item{the signal strength in dBm, \textit{RSSI}.}
\end{itemize}

APs do not have geographical coordinates attached, but their position tend to be fixed. The geographical position of APs is estimated the procedure described in the Supplementary Information, which used participants' sequences of GPS scans to obtain \textit{APs} locations and remove mobile \textit{APs}. Then, we clustered geo-localised \textit{APs} to ``locations" using a graph-based approach. With our definition, a ``location'' is a connected component in the graph $G_d$, where a link exists between two \textit{APs} if their distance is smaller than a threshold $d$ (see \cite{1609.03526}, SI for more details). Here, we present results obtained for $d = 2~\mathrm{m}$. However, results are robust with respect to the choice of the threshold (see also \cite{1609.03526}). 

Throughout the experiment, participants' devices scanned for WiFi every $\Delta t$ seconds. The median time between scans is between $\Delta t_M=16~\mathrm{s}$ and $\Delta t_M<60~\mathrm{s}$ for 90\% of the population (see also \cite{1609.03526}, SI). Data was temporally aggregated in bins of length $\Delta t=60~\mathrm{s}$, since we focus here on the $\emph{pauses}$ between moves. If a participant visits more than one location within a timebin, we assign the location in which they spent the most time to that bin. Given our definition of location and the given time-binning, the median daily time coverage (the fraction of minutes/day that an individual's position is known, where the median is taken across all days) is included between $0.6$ and $0.98$ for 90\% of the population.

\subsection*{Model selection}

\noindent The best model is selected using Akaike weights~\cite{wagenmakers2004aic}.
First, we determine the best fit parameters for each of the models via Nelder-Mead numerical Likelihood maximisation~\cite{nelder1965simplex} (maximisation is considered to fail if convergence with tolerance $t=0.0001$ is not reached after $200 \cdot N$ iterations, where $N$ is the length of the data). 
For each model $m$, we compute the Akaike Information Criterion: 

\begin{equation}
AIC_m=-2\log{L_m}+2V_m+ \frac{2V_m(V_m+1)}{n-V_m-1}
\end{equation}

\noindent where $L_m$ is the maximum likelihood for the candidate model $m$, $V_m$ is the number of free parameters in the model, and $n$ is the sample size. 
The $AIC$ reaches its minimum value $AIC_{min}$ for the model that minimises the expected information loss. Thus, AIC rewards descriptive accuracy via the maximum likelihood and penalises models with large number of parameters. 

The Akaike $w_m(AIC)$ weight of a model $m$ corresponds to its relative likelihood with respect to a set of possible models. Measuring the Akaike weights allows us to compare the descriptive power of several models. 
 
\begin{equation}
w_m(AIC)=\dfrac{e^{-\frac{1}{2}(AIC_m-AIC_{min})}}{\sum\limits_{k=1}^K e^{-\frac{1}{2}(AIC_k-AIC_{min})}}
\end{equation}

\noindent For all distributions considered in this paper, we found one model $m*$ such that $w_m*\sim 1$ (which implies all the other models have Akaike weight very close to 0).

\subsection*{Figures}
All figures were generated using Matplotlib \cite{hunter2007matplotlib} package (version 1.5.3) for Python.

\section*{Related work}

We present here more detailed analysis of the literature discussed in the paper.

\def\arraystretch{2}

\begin{longtable}{||p{55pt} p{30pt} p{40pt} p{40pt} p{40pt} p{100pt} p{45pt} p{65pt} ||}
\caption{\textbf{Distribution of waiting times and displacements: a comparison of over 30 datasets on human mobility } The table reports for each dataset: the reference to the journal article/book where the study was published, the type of data (LBSN stands for Location Based Social Networks, CDR for Call Detail Record), the number of individuals (or vehicles in the case of car/taxi data) involved in the data collection, the duration of the data collection (M $\rightarrow$ months, Y $\rightarrow$ years, D $\rightarrow$ days, W $\rightarrow$ weeks), the minimum and maximum length of spatial displacements, the shape of the probability distribution of displacements with the corresponding parameters, the temporal sampling, the shape of the distribution of waiting times with the corresponding parameters. Power-law (T), indicates a truncated power-law. The table can also be found at \url{http://lauraalessandretti.weebly.com/plosmobilityreview.html}}\label{theTable}\\
\endfirsthead
\\

\hline
\textbf{} & \textbf{Data type} & \textbf{N} & \textbf{Dur.} & \textbf{Range \newline $\Delta x$} & \textbf{$P(\Delta x$)} & \textbf{Sampling $\delta t$} & \textbf{$P(\Delta t$)}\\
\hline
\endhead 
\hline
\textbf{} & \textbf{Data type} & \textbf{N} & \textbf{Dur.} & \textbf{Range \newline $\Delta x$} & \textbf{$P(\Delta x$)} & \textbf{Sampling $\delta t$} & \textbf{$P(\Delta t$)}\\
\hline
\cite{song2010modelling} (D1) & CDR & $3.0 \cdot10^{6}$ 

& $1$ Y & $1~\mathrm{km}$ \newline $100~\mathrm{km}$ & \textbf{power-law (T)} \newline $\beta$=1.55 & uneven &  
\\
\hline

\cite{song2010modelling} (D2) & CDR & $10^{3}$ & $2$ W  & $1~\mathrm{km}$ \newline $100~\mathrm{km}$ &  &   $1$ h  & \textbf{power-law (T)} \newline $\beta$=1.80  
\\
\hline

\cite{gonzalez2008understanding} (D1) & CDR & $10^{5}$ & $6$ M  & $1~\mathrm{km}$ \newline $1000~\mathrm{km}$ & \textbf{power-law (T)} \newline $\beta$=1.75 & uneven   & 
\\
\hline
\cite{gonzalez2008understanding} (D2) & CDR & $206$ & $1$ W  & $1~\mathrm{km}$ \newline $500~\mathrm{km}$ & \textbf{power-law (T)} \newline $\beta$=1.75 & $2$ h   & 
\\
\hline

\cite{brockmann2006scaling} & Bills records & $4.6 \cdot10^{5}$\newline bills & $1.39$ Y  & $100~\mathrm{m}$\newline $3200~\mathrm{km}$ & 10$\leqslant\Delta x\leqslant 3200~\mathrm{km}$ \newline \textbf{power-law} \newline $\beta$=1.59 & uneven   & 
\\
\hline
\cite{wang2014correlations} (Geolife) & GPS & $32$ & $3.42$ Y  & $10~\mathrm{m}$ \newline $10000~\mathrm{km}$ & 0.01 $\leqslant \Delta x \leqslant$ $10~\mathrm{km}$ \newline  \textbf{power-law} \newline $\beta_{0}$=1.25 \newline \newline 10 $< \Delta x \leqslant$ $10000~\mathrm{km}$ \newline  \textbf{power-law} \newline  $\beta_{1}$=1.90 & $2$ min  & \textbf{power-law}  $\beta$=1.98
\\
\hline

\cite{zhao2015explaining} \newline (Nokia) & GPS & $200$ & $1.50$ Y  & $100~\mathrm{m}$ \newline $10~\mathrm{km}$ & \textbf{power-law (T)} \newline  $\beta$=1.39 & $10$ sec &  
\\
\hline
\cite{zhao2015explaining} (Geolife) & GPS & $182$ & $5.00$ Y  & $100~\mathrm{m}$ \newline $10~\mathrm{km}$ & \textbf{power-law (T)} \newline $\beta$=1.57 & $1-5$ sec   & 
\\
\hline
\cite{rhee2011levy} \\ (5 datasets) & GPS  & $101$ & $5$ M  & $10~\mathrm{m}$ \newline $10~\mathrm{km}$ & \textbf{power-law (T)} \newline $\beta$=[1.35-1.82] & $10$ sec  & \textbf{power-law (T)} \newline $\beta$=[1.45-2.68]
\\
\hline
\cite{jiang2009characterizing} & Taxi (GPS) & $50$ & $6$ M  & $1~\mathrm{Km}$ \newline $100~\mathrm{km}$ & 3 $\leqslant \Delta x \leqslant$ $23~\mathrm{km}$ \newline  \textbf{power-law} \newline $\beta_{0}$=2.50  \newline \newline 23 $< \Delta x \leqslant$ $100~\mathrm{km}$ \newline  \textbf{power-law} \newline $\beta_{1}$=4.60 & $10$ sec    & 
\\
\hline
\cite{liu2012understanding} & Taxi (GPS) & $6.6 \cdot10^{3}$ & $1$ W  & $1~\mathrm{km}$ \newline $100~\mathrm{km}$ & \textbf{power-law (T)} \newline $\beta$=1.20 & $10$ sec   & 
\\
\hline
\cite{beiro2016predicting} & Flickr & $4.0 \cdot10^{4}$ &  & $1~\mathrm{km}$ \newline $10000~\mathrm{km}$ & \textbf{power-law (T)}   & uneven   & 
\\
\hline
\cite{cheng2011exploring} & LBSN & $2.2 \cdot10^{5}$ & $4$ M  & $1~\mathrm{km}$ \newline $500~\mathrm{km}$ & \textbf{power-law}\newline $\beta$=1.88 & uneven   & 
\\
\hline
\cite{hawelka2014geo} & Twitter & $1.3 \cdot10^{7}$ & $1$ Y & $1~\mathrm{km}$ \newline $100~\mathrm{km}$ & \textbf{power-law} \newline $\beta$=1.62 & uneven   & 
\\
\hline

\cite{noulas2012tale} & LBSN & $9.2 \cdot10^{5}$ & $6$ M  & $1~\mathrm{km}$ \newline $20000~\mathrm{km}$ & \textbf{power-law} \newline $\beta$=1.50 & uneven &  
\\
\hline
\cite{noulas2012tale} (intracity)& LBSN & $9.2 \cdot10^{5}$ & $6$ M  & $10~\mathrm{m}$ \newline $100~\mathrm{km}$ & \textbf{power-law \newline (``poor'')\cite{noulas2012tale}} \newline $\beta$=4.67 & uneven &  
\\
\hline
\cite{wu2014intra} & LBSN & $2.6 \cdot10^{5}$& $1$ Y & $10~\mathrm{m}$ \newline $50~\mathrm{km}$ & \textbf{exponential} \newline $\lambda=$0.179 & uneven   & 
\\
\hline
\cite{liu2014uncovering} & LBSN & $5.2 \cdot10^{5}$ & $1$ Y & $1~\mathrm{km}$ \newline $4000~\mathrm{km}$ & \textbf{exponential} \newline $\lambda$=0.003 & uneven   & 
\\
\hline
\cite{jurdak2015understanding} & Twitter & $1.6 \cdot10^{5}$ & $8$ M  & $10~\mathrm{m}$ \newline $4000~\mathrm{km}$ & 0.01 $\leqslant \Delta x \leqslant$ $0.1~\mathrm{km}$ \newline  \textbf{exponential}\newline $\lambda$=0.073\newline \newline 0.1 $< \Delta x \leqslant$ $100~\mathrm{km}$ \newline  \textbf{Stretched \newline power-law} \newline $\beta_{1}$=0.45 \newline  \newline 100 $< \Delta x \leqslant$ $4000~\mathrm{km}$ \newline  \textbf{power-law}  \newline $\beta_{2}$=1.32  \newline  & uneven   & 
\\
\hline

\cite{liu2015crossover} & Taxi (GPS) & $803$ & $1.25$ Y  & $1~\mathrm{km}$ \newline $100~\mathrm{km}$ & $ \Delta x \leqslant$ $15~\mathrm{km}$ \newline  \textbf{exponential} \newline $\lambda$=0.36\newline \newline  15 $< \Delta x \leqslant$ $100~\mathrm{km}$ \newline  \textbf{power-law}   \newline $\beta$=3.66 & $30$ sec   & 
\\
\hline

\cite{liang2012scaling} (D1) & Taxi (GPS) & $10^{4}$ & $3$ M  & $1~\mathrm{km}$ \newline $100~\mathrm{km}$ & 1 $\leqslant \Delta x \leqslant$ $20~\mathrm{km}$ \newline \textbf{exponential} \newline  $\lambda_{0}$=0.23 \newline \newline 20 $< \Delta x \leqslant$ $100~\mathrm{km}$ \newline  \textbf{exponential} \newline  $\lambda_{1}$=0.17\newline  & $1$ min    & 
\\

\hline

\cite{liang2012scaling}  (D2) & Taxi (GPS) & $10^{4}$ & $2$ M & $1~\mathrm{km}$ \newline $100~\mathrm{km}$ & 1 $\leqslant \Delta x \leqslant$ $20~\mathrm{km}$ \newline  \textbf{exponential} \newline $\lambda_{0}$=0.24\newline \newline  20 $< \Delta x \leqslant$ $100~\mathrm{km}$ \newline  \textbf{exponential} \newline  $\lambda_{1}$=0.18& $1$ min    & 
\\
\hline
\cite{gong2016inferring} & Taxi (GPS) & $6.6 \cdot10^{3}$ & $1$ W  & $2~\mathrm{km}$ \newline $20~\mathrm{km}$ & \textbf{exponential} \newline $\lambda$=[0.072-0.252] & $10$ sec  &  
\\
\hline

\cite{zhao2015automatic} \newline (3 datasets) & Taxi (GPS) &$10^{4}$ & $1$ M & $600~\mathrm{m}$ \newline $10~\mathrm{km}$ & \textbf{exponential}  & $[9-177]$ s  & \textbf{power-law} 
\\

\hline
\cite{wang2015comparative} \newline (6 datasets) & Taxi (GPS) & $3.0 \cdot10^{4}$ & [$1$ M-$2$ Y]  & $1~\mathrm{km}$ \newline $100~\mathrm{km}$ & \textbf{log-normal}\newline $\mu$=[0.77-1.32], \newline $\sigma$=[0.67-0.87] & [24 - 116] s    &  
\\

\hline

\cite{tang2015uncovering} & Taxi (GPS) & $1.1 \cdot10^{3}$ & $6$ M  & $100~\mathrm{m}$ \newline $30~\mathrm{km}$ & \textbf{log-normal} \newline $\mu$=0.38, \newline $\sigma$=0.48 & $30$ sec    & 
\\

\hline

\cite{schneider2013unravelling} & Surveys & $10^{4}$ & $1$ Y &  &    & self-reported & \textbf{power-law (T)} \newline $\beta$=0.49
\\

\hline

\cite{gallotti2016stochastic} & Private Cars (GPS) & $7.8  \cdot10^{5}$ & $1$ M & $1~\mathrm{km}$ \newline $500~\mathrm{km}$ & \textbf{superimposition 
\newline Poisson}   & $10$ sec  &  $\Delta t \leqslant $4h \newline  \textbf{power-law} \newline $\beta$=1.03 \newline \newline 4 $\leqslant \Delta t \leqslant $200h \newline  \textbf{log-normal}  \newline $\mu$=1.60,\newline $\sigma$=1.60\\

\hline

\cite{bazzani2010statistical} & Private Cars (GPS) & $3.5 \cdot10^{4}$& $1$ M & $300~\mathrm{m}$ \newline $100~\mathrm{km}$ & \textbf{polynomial}  & $10$ sec  & \textbf{power-law} \newline $\beta$=0.97
\\

\hline

\cite{deville2016scaling} (D1) & CDR & $1.3  \cdot10^{6}$ & $1$ M & $1~\mathrm{km}$ \newline $200~\mathrm{km}$ & \textbf{power-law}\newline $\beta=2.02$ & uneven  &   \\
\hline

\cite{deville2016scaling} (D2) & CDR & $6  \cdot10^{6}$ & $1$ Y & $1~\mathrm{km}$ \newline $500~\mathrm{km}$ & \textbf{power-law} \newline $\beta=1.75$ & uneven   &  \\
\hline

\cite{deville2016scaling} (D3) & CDR & & $4$ Y & $1~\mathrm{km}$ \newline $100~\mathrm{km}$ & \textbf{power-law} \newline $\beta=1.80$  & uneven  & \\

\hline
\cite{roth2011structure} & Travel cards & $2.0 \cdot10^{6}$ & $1$ W & $100~\mathrm{m}$ \newline $50~\mathrm{km}$ & \textbf{negative binomial}\newline $\mu$=9.28,\newline $\sigma$=5.83 & uneven &  
\\
\hline

\cite{yan2013diversity} & Travel \newline Diaries & $230$ & $1.5$ M  & $1~\mathrm{km}$ \newline $400~\mathrm{km}$ & \textbf{power-law (T)} \newline $\beta$=1.05 & self-reported &  
\\
\hline

\cite{riccardo2012towards} & Private Cars (GPS) & $7.5 \cdot10^{4}$ & $1$ M & $10~\mathrm{m}$ \newline $500~\mathrm{km}$ & 0.01 $\leqslant \Delta x \leqslant$ $20~\mathrm{km}$ \newline  \textbf{exponential} \newline \newline 20 $< \Delta x \leqslant$ $150~\mathrm{km}$ \newline \textbf{power-law}  \newline $\beta_{1}$=3.30 & $30$ sec  & $\Delta t \leqslant $3h \newline  \textbf{exponential}\newline $\lambda$=1.02
\\

\hline

\cite{yao2016study} & Taxi (GPS) & \newline  & $1$ D  & $200~\mathrm{m}$ \newline $1000~\mathrm{km}$ & \textbf{power-law}\newline $\beta$=2.70 &  &  \\
\hline
\end{longtable}

\captionsetup{margin={-0in,0in}}

\clearpage

{\begin{center}

\Large\bf{Supporting Information \\
Multi-scale spatio-temporal analysis of human mobility
}

\end{center}}

\beginsupplement
\section{Data pre-processing}

\subsection*{Determining routers' locations}
We determine the routers' locations using the approach described in \cite{sapiezynski2015opportunities} with a slight modification. The original method used only GPS location estimations calculated at the same second as a corresponding WiFi scan. Here, we consider all location estimations from Android Location API, including network based estimations.
Additionally, we relax the same-second requirement as follows. In the spatio-temporal trace of each user we identify periods from time $t_0$ to time $t_N$ where the user was stationary, also referred to as stop locations. This means that the distance between the user's location at $t_0$ and $t_N$ is below $d$ meters, and that there exist a location estimation between $t_0$ and $t_N$ at least every $n$ seconds. Also, it implies that each location estimation within the stop location is within $d$ from the user's location at $t_N$. At $t_N$, the individual stop-location changes. We select $n$ as 305 seconds, thus requiring no missing data, since the sampling period of GPS location in the experiment is approximately 300 seconds. We select $d$ as 30 meters, a safe range compared to the typical GPS errors, thus requiring that the user remains in the same location within the resolution of a building. After identifying these stop locations, we assign the geometric median position of estimations to all routers scanned in these periods.

\section{Robustness of results}

\subsection*{Results of the model selection}

\begin{figure}[h!]
\centering
\includegraphics[width=\textwidth]{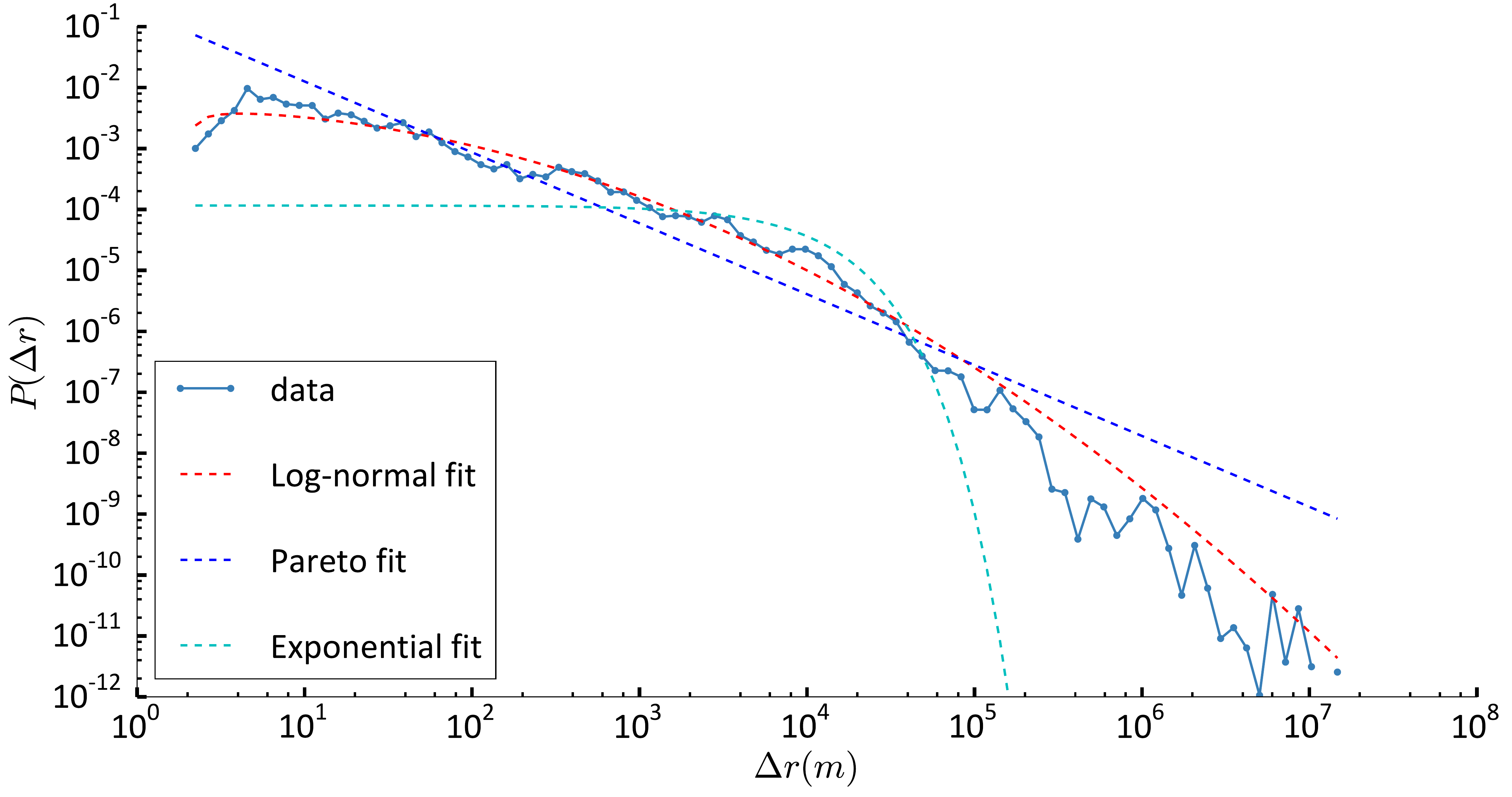}
\caption{\textbf{Distribution of displacements: comparison of three models}. Blue dotted line: data. Red dashed line: Maximum likelihood Log-normal fit. Blue dashed line: Maximum likelihood Pareto fit. Light blue dashed line: Maximum likelihood Exponential fit.}
\label{disp}
\end{figure}

\begin{figure}[h!]
\centering
\includegraphics[width=\textwidth]{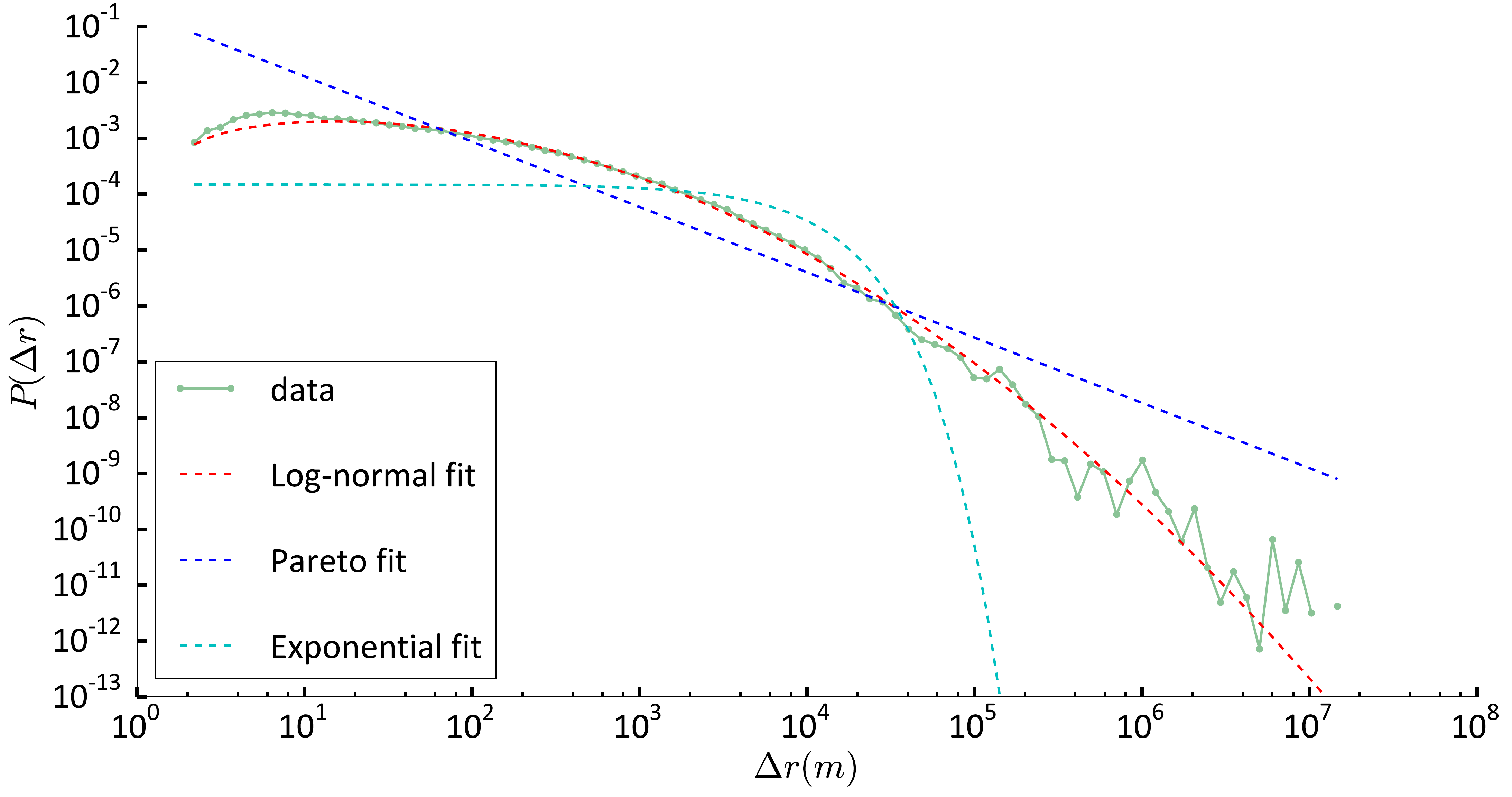}
\caption{\textbf{Distribution of waiting times: comparison of three models}. Yellow dotted line: data. Red dashed line: Maximum likelihood Log-normal fit. Blue dashed line: Maximum likelihood Pareto fit. Light blue dashed line: Maximum likelihood Exponential fit.}
\label{wt}
\end{figure}

\begin{figure}[h!]
\centering
\includegraphics[width=\textwidth]{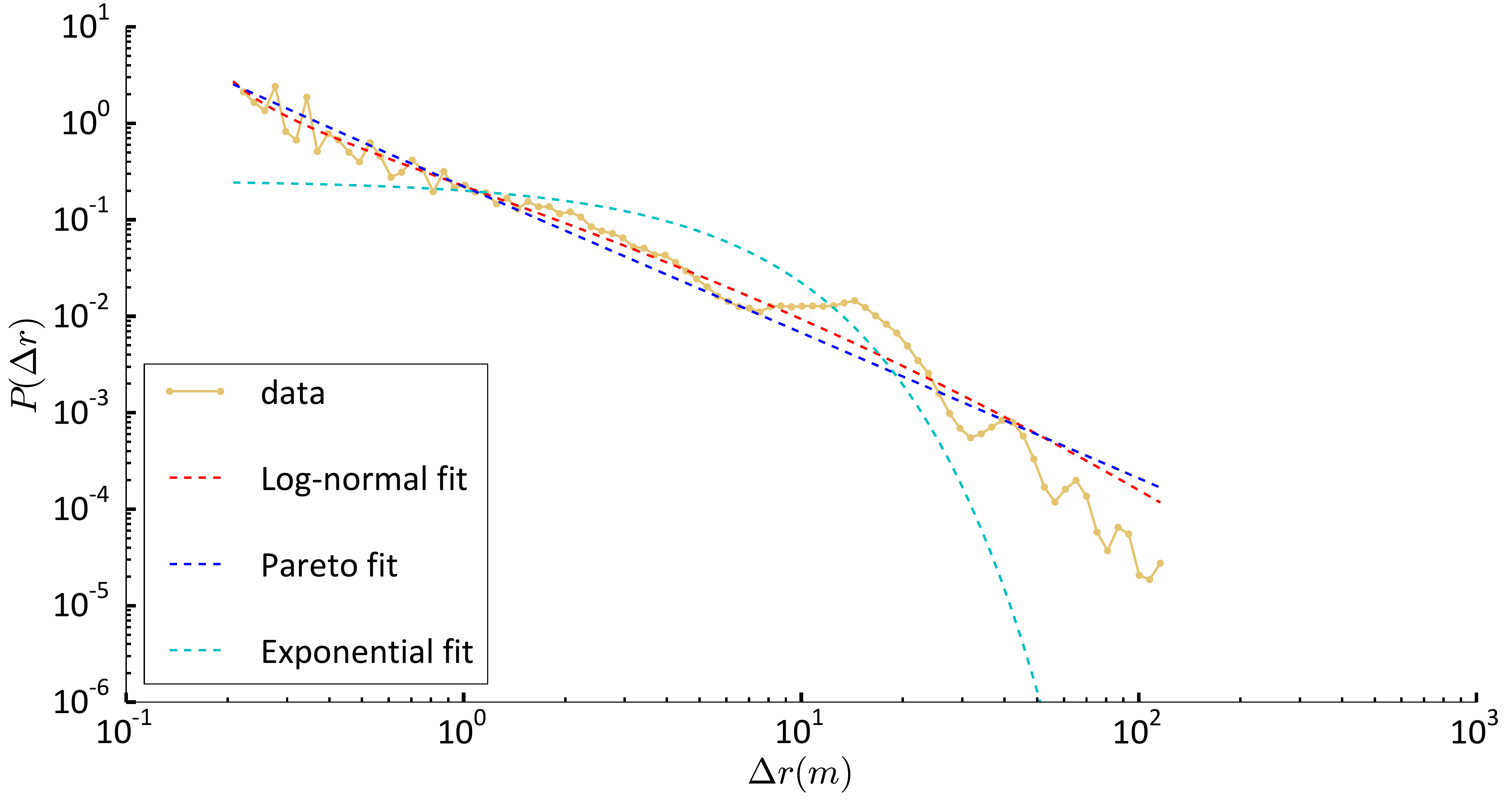}
\caption{\textbf{Distribution of displacements between discoveries: comparison of three models}. Green dotted line: data. Red dashed line: Maximum likelihood Log-normal fit. Blue dashed line: Maximum likelihood Pareto fit. Light blue dashed line: Maximum likelihood Exponential fit.}
\label{disc}
\end{figure}

The selection of the log-normal distribution as the best model among the exponential, the log-normal and the Pareto distribution is made using the Akaike Information Criterion (AIC) weights. In tables \ref{TableDisplacements},\ref{TableWT},\ref{TableDiscoveries} we report the AIC weights values for the four models considered as well as the Akaike information Criterion (AIC), the Bayesian Information Criterion (BIC) weights, the Residual Sum of Squares (RSS). These metrics provide additional information on the goodness-of-fit. In figures \ref{disp}, \ref{wt}, \ref{disc}, we show the results of the fit with the three distributions considered.

\def\arraystretch{1.5}

\begin{table}[h!]
\begin{tabular}{lrrrr}
\hline
         &         AIC &   AIC weights &   BIC weights &   RSS  \\
\hline
 expon   &      2.1e+07 &             0 &                 0 & 3.1e-11  \\
 lognorm &      1.9e+07 &             1 &                 1 & 2.9e-11 \\
 pareto  &      2.0e+07  &             0 &                 0 & 2.8e-11 \\
\hline
\end{tabular}
\caption{\textbf{Distribution of displacements: model selection}. For the three distributions considered, the table reports the Akaike Information Criterion (AIC), the AIC weights (see Model selection section), the Bayesian Information Criterion (BIC) and the residual sum of squares (RSS).}
\label{TableDisplacements}
\end{table}

\begin{table}[h!]
\begin{tabular}{lrrrr}
\hline
         &         AIC &   AIC weights &   BIC weights &   RSS  \\
\hline
 expon   &      4.62e+06 &             0 &                 0 & 0.061 \\
 lognorm &      3.68e+06 &             1 &                 1 & 0.026 \\
 pareto  &      3.79e+06 &             0 &                 0 & 0.025 \\
\hline
\end{tabular}
\caption{\textbf{Distribution of waiting times: model selection}. For the three distributions considered, the table reports the Akaike Information Criterion (AIC), the AIC weights (see Model selection section), the Bayesian Information Criterion (BIC) and the residual sum of squares (RSS).}
\label{TableWT}
\end{table}
\begin{table}[h!]
\begin{tabular}{lrrrr}
\hline
         &         AIC &   AIC weights &   BIC weights &   RSS  \\
\hline
 lognorm &      2.7e+07 &             1 &                 1 & 3.0e-11 \\
 pareto  &      2.9e+07  &             0 &                 0 & 2.8e-11 \\
 expon   &      3.0e+07 &             0 &                 0 & 3.1e-11 \\
\hline
\end{tabular}
\caption{\textbf{Distribution of displacements between discoveries: model selection}. For the three distributions considered, the table reports the Akaike Information Criterion (AIC), the AIC weights (see Model selection section), the Bayesian Information Criterion (BIC) and the residual sum of squares (RSS).}
\label{TableDiscoveries}
\end{table}
\subsection*{Bootstrapping}

\begin{figure}[h!]
\centering
\includegraphics[width=\textwidth]{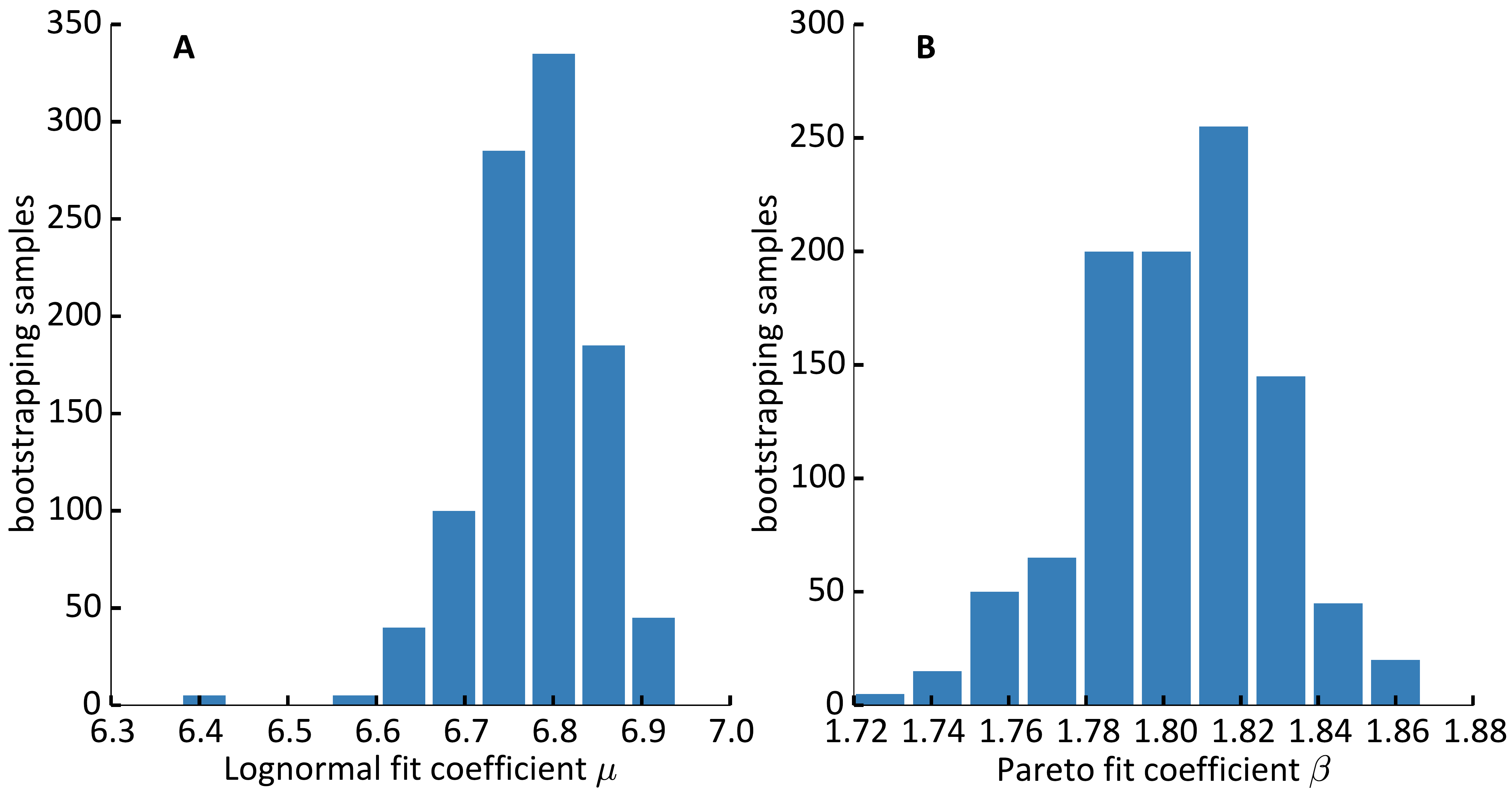}
\caption{\textbf{Displacements: distribution of parameters found by bootstrapping}. \textbf{A)}The distribution over 1000 bootstrapping samples of the log-normal fit coefficient $\mu$, characterising the aggregated distribution of displacements. \textbf{B)}The distribution over 1000 bootstrapping samples of the Pareto fit coefficient $\beta$, characterising the tail of the aggregated distribution of displacements. Samples include $100$ randomly selected individuals. }
\label{B1}
\end{figure}

\begin{figure}[h!]
\centering
\includegraphics[width=\textwidth]{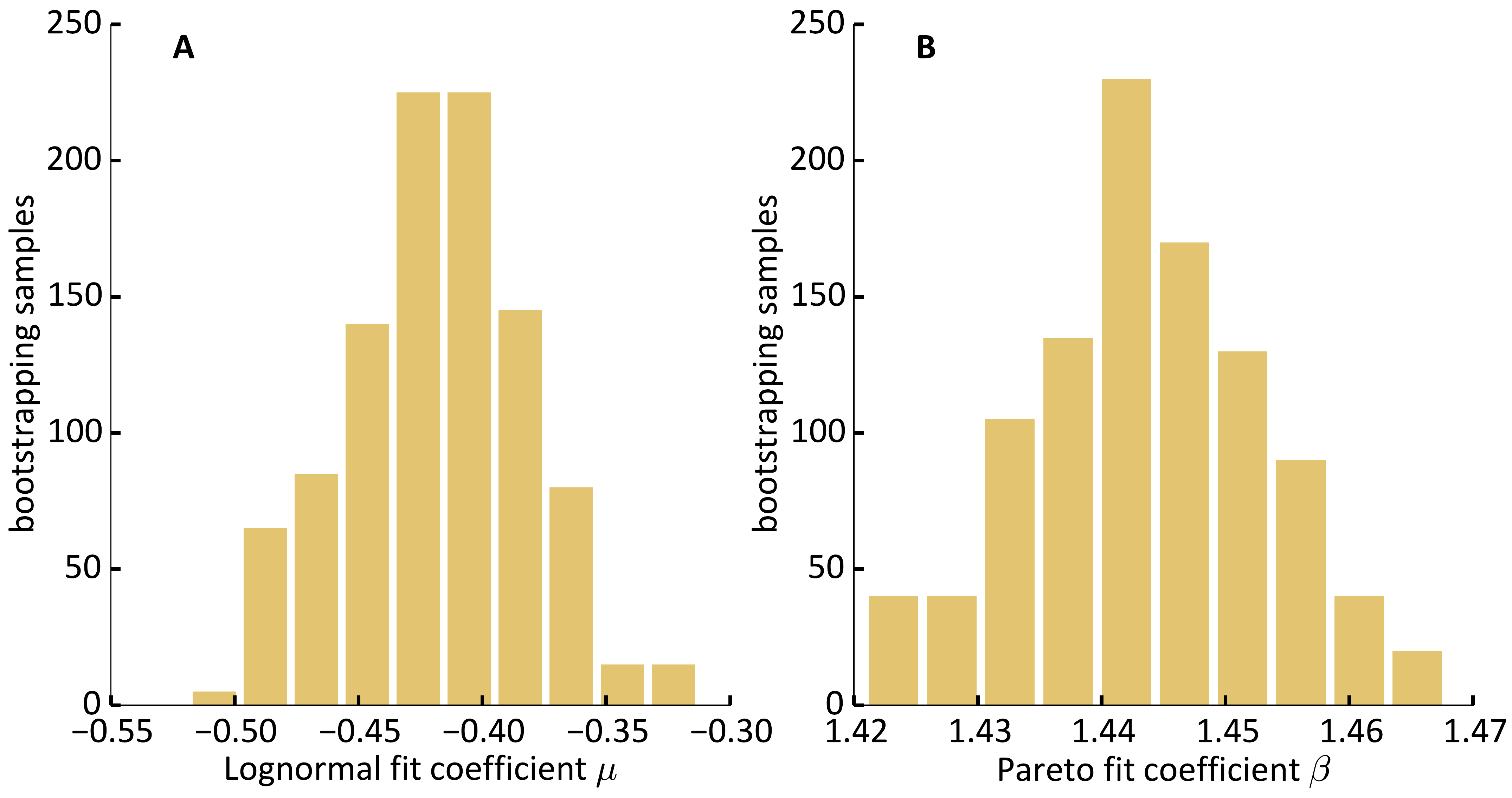}
\caption{\textbf{Waiting Times: distribution of parameters found by bootstrapping}. \textbf{A)}The distribution over 1000 bootstrapping samples of the log-normal fit coefficient $\mu$, characterising the aggregated distribution of waiting times. \textbf{B)}The distribution over 1000 bootstrapping samples of the Pareto fit coefficient $\beta$, characterising the tail of the aggregated distribution of waiting times. Samples include $100$ randomly selected individuals.}
\label{B2}
\end{figure}

\begin{figure}[h!]
\centering
\includegraphics[width=\textwidth]{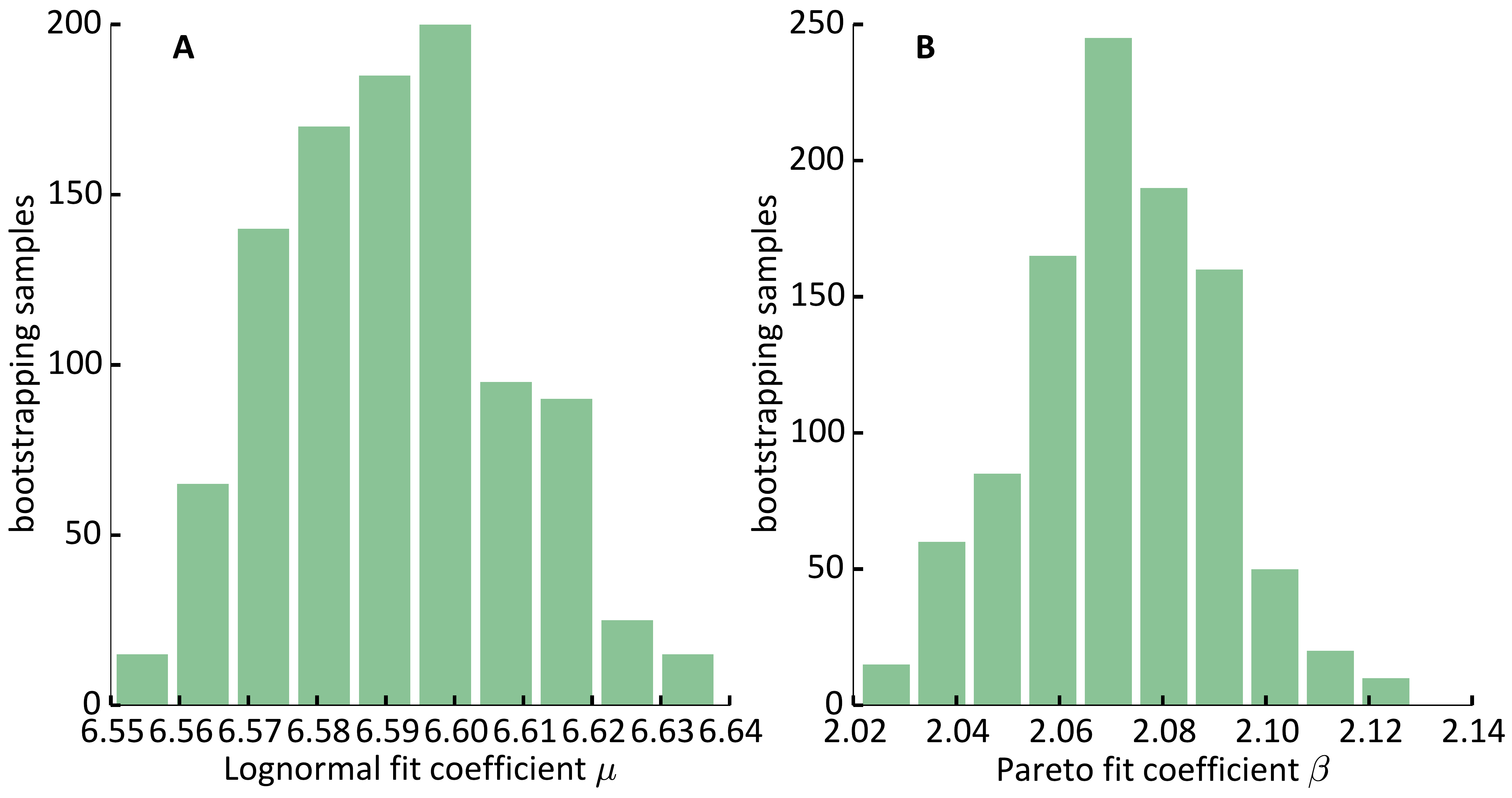}
\caption{\textbf{Displacements between discoveries: distribution of parameters found by bootstrapping}. \textbf{A)}The distribution over 1000 bootstrapping samples of the log-normal fit coefficient $\mu$, characterising the aggregated distribution of displacements between discoveries. \textbf{B)}The distribution over 1000 bootstrapping samples of the Pareto fit coefficient $\beta$, characterising the tail of the aggregated distribution of displacements between discoveries. Samples include $100$ randomly selected individuals.}
\label{B3}
\end{figure}

By bootstrapping data $1000$ times for samples of $100$ and $200$ individuals, we find that for all groups the aggregated distributions of displacements and waiting times are best described by the same models found for the entire dataset.\\
Here, we report the distribution of parameters found for the distribution of displacements (Fig \ref{B1}), waiting times (Fig \ref{B2}), and displacements between discoveries (Fig \ref{B3}), in the case of samples of $100$ individuals. 

\subsection*{Sensitivity to the definition of pausing}

\begin{figure}[h!]
\centering
\includegraphics[width=\textwidth]{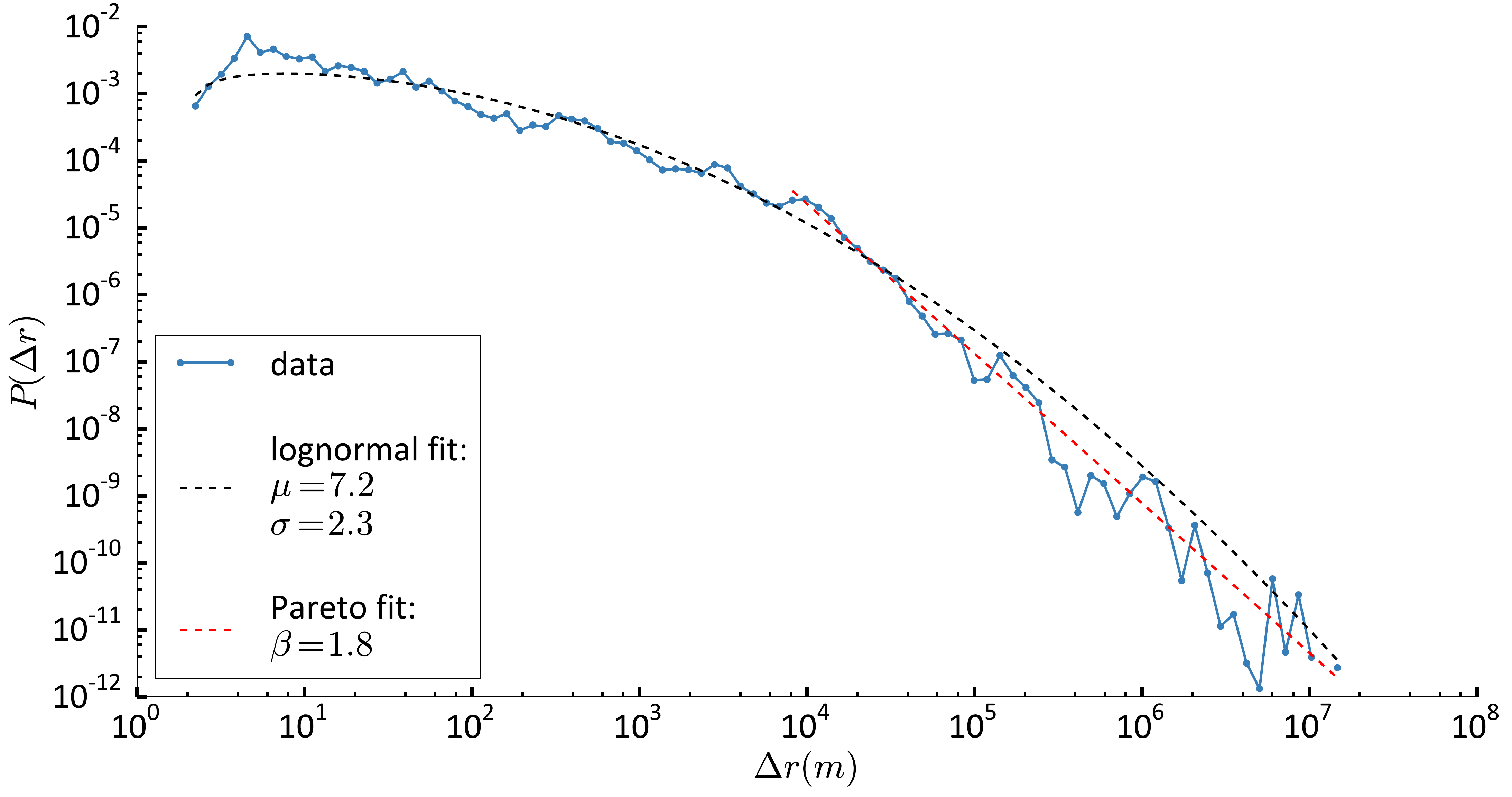}
\caption{\textbf{Distribution of displacements for pausing P=15 minutes}. Blue dotted line: data. Black dashed line: Log-normal fit with characteristic parameter $\mu$ and $\sigma$. Red dashed line: Pareto fit with characteristic parameter $\beta$ for $\Delta r>7420~\mathrm{m}$.}
\label{P=15}
\end{figure}

\begin{figure}[h!]
\centering
\includegraphics[width=\textwidth]{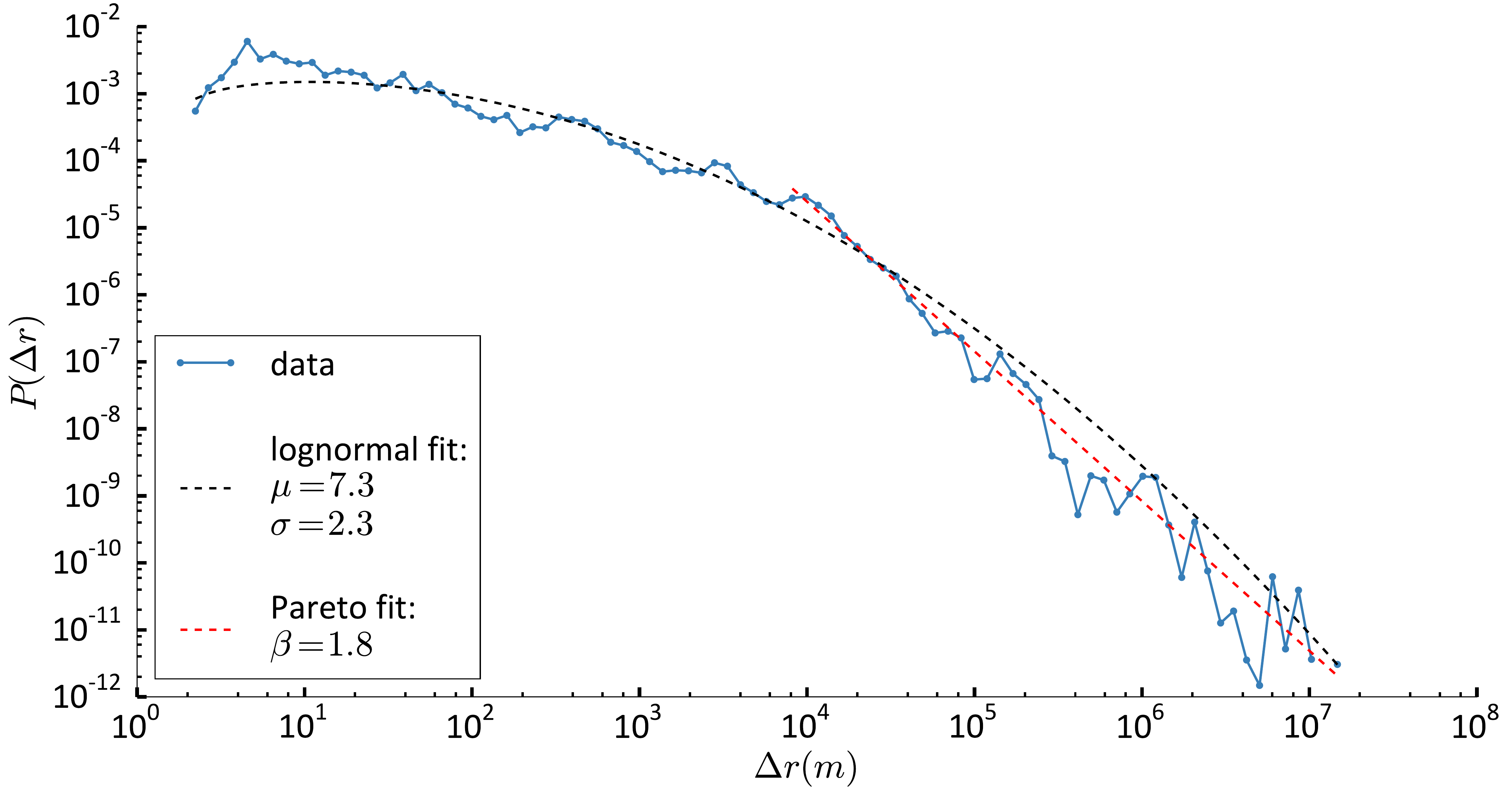}
\caption{\textbf{Distribution of displacements for pausing P=20 minutes}. Blue dotted line: data. Black dashed line: Log-normal fit with characteristic parameter $\mu$ and $\sigma$. Red dashed line: Pareto fit with characteristic parameter $\beta$  for $\Delta r>7420~\mathrm{m}$.}
\label{P=20}
\end{figure}

\begin{figure}[h!]
\centering
\includegraphics[width=\textwidth]{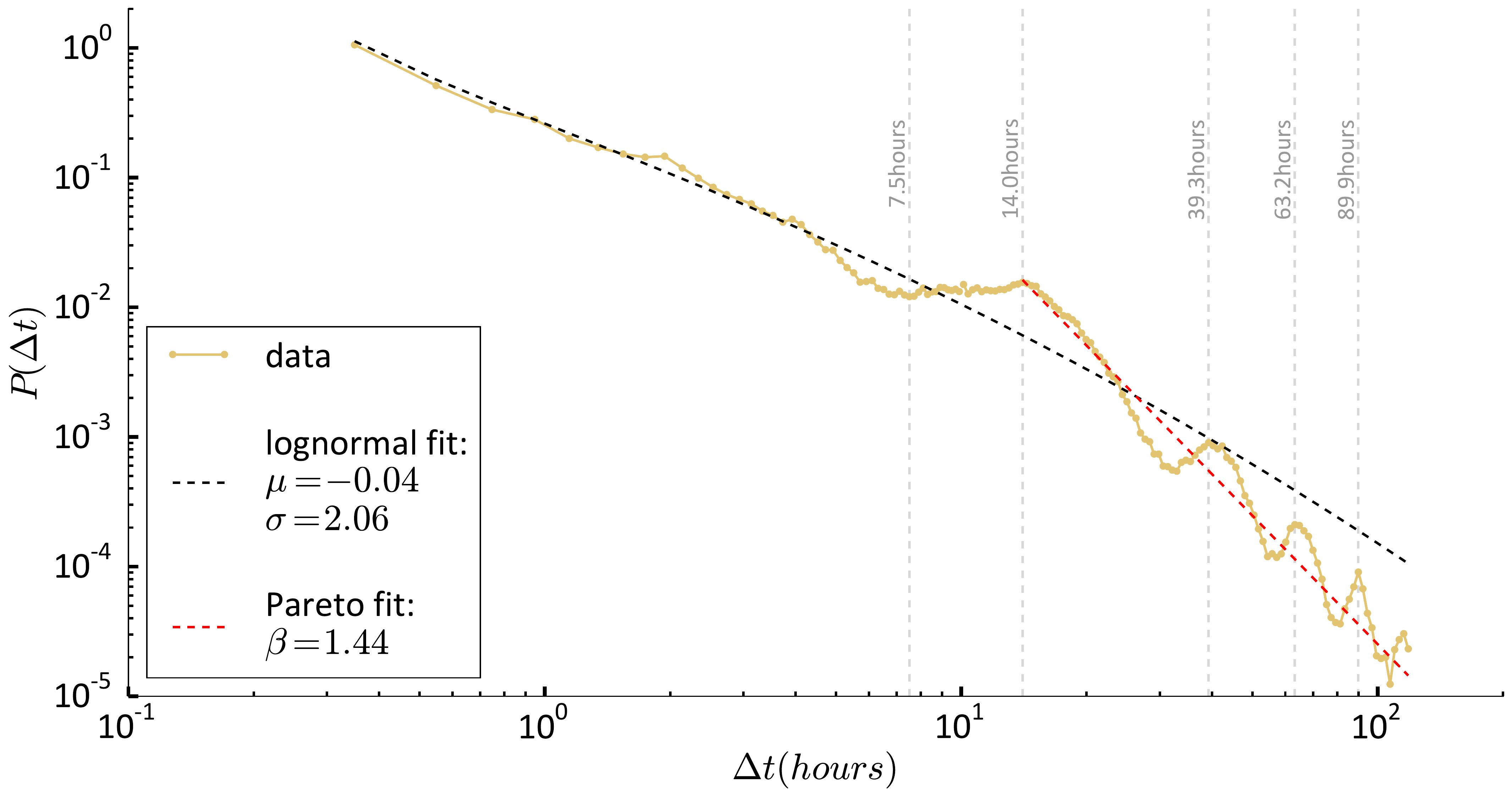}
\caption{\textbf{Distribution of waiting times for pausing P=15 minutes}. Yellow dotted line: data. Black dashed line: Log-normal fit with characteristic parameter $\mu$ and $\sigma$. Red dashed line: Pareto fit with characteristic parameter $\beta$ for $\Delta t>13 h$.}
\label{P=15_2}
\end{figure}

\begin{figure}[h!]
\centering
\includegraphics[width=\textwidth]{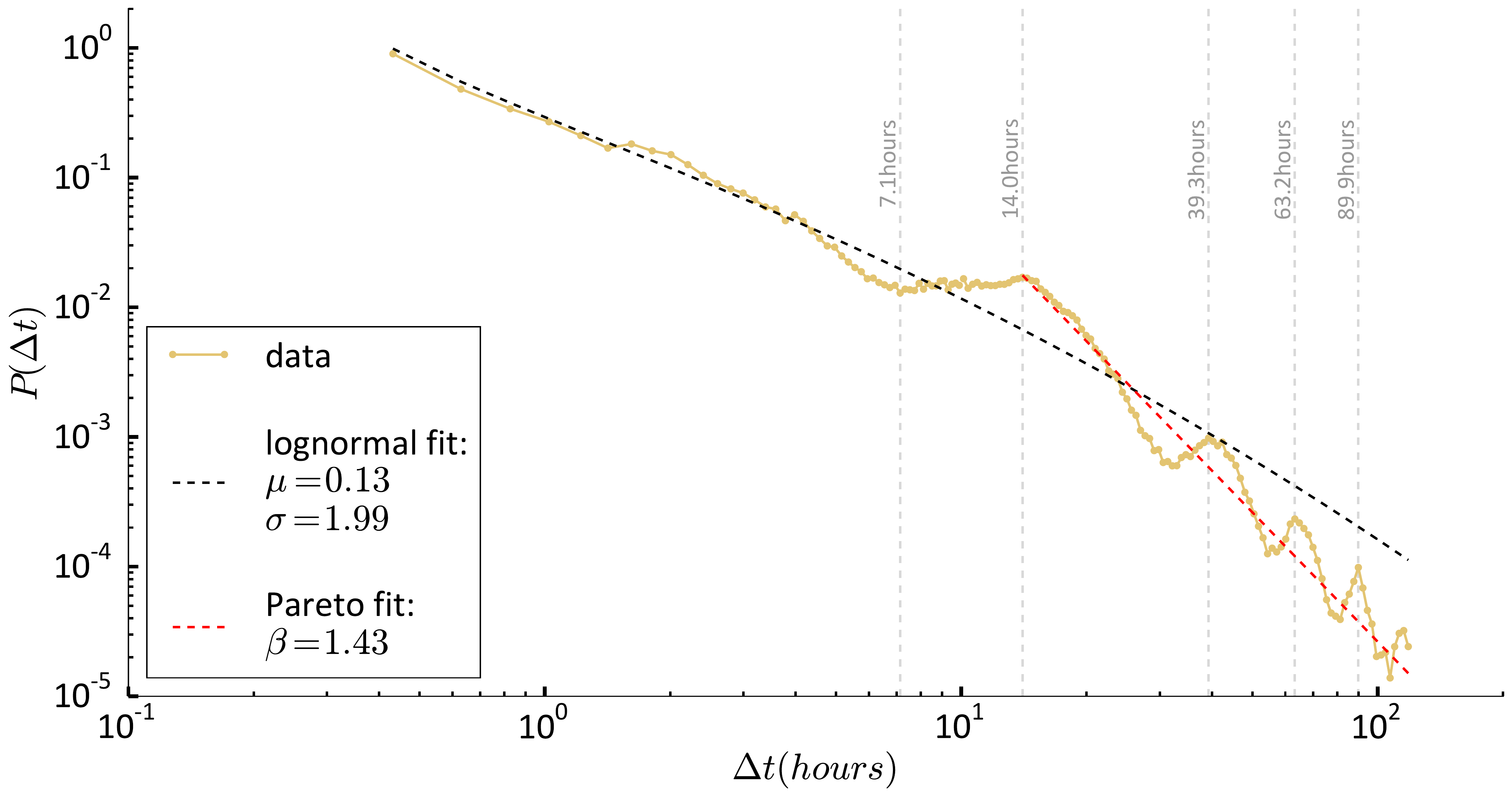}
\caption{\textbf{Distribution of displacements for pausing P=20 minutes}. Yellow dotted line: data. Black dashed line: Log-normal fit with characteristic parameter $\mu$ and $\sigma$. Red dashed line: Pareto Fit with characteristic parameter $\beta$ for $\Delta t>13~\mathrm{h}$.}
\label{P=20_2}
\end{figure}

The distribution of displacements is robust with respect to the definition of \emph{pausing}. The results reported in the main text refer to pauses longer than $P=10$ minutes. Both for $P=15$ minutes and $P=20$ minutes, the distribution of displacements is best described by a log-normal model when the entire distribution is taken into account, and by a Pareto distribution, when only long distances are considered (see Figures~\ref{P=15} and~\ref{P=20}). The same results hold for the distributions of waiting times (see Figures~\ref{P=15_2} and~\ref{P=20_2})

\subsection*{Interpretation of the shift and scale parameters}

The shift and scale parameters are necessary to account for the fact that, in the cases considered, the lower bound of the distributions support is controlled by the data minimal resolution.

For example, the log-normal distribution of a random variable $x$ is defined for $x \in (0,\infty)$. In our case the fit is performed for a shifted distribution, with $x \in (x_0,\infty)$, where $x_0$ is the data minimal resolution. This reflects the fact that the reason why there are no data points for $x<x_0$ is not low probability but lack of information within this range (or in some cases it's due to the choice of fitting only the tail of the distribution).

Similarly, the Pareto distribution is defined for $x \in (1, \infty)$. The shift $x_0$ and the scale parameter $s$ allow instead to consider $x \in (s+x_0,\infty)$, where $s+x_0$ is the minimum data point considered. Values of the shift and scale parameters could be set to fit the minimal resolution. However, in our case $x_0$ and $s$ are additional parameters of the model. We have verified that the values recovered by the fitting algorithm are consistent with those expected. 

We report in table \ref{TableShiftScale} the values of the shift $s$ and scale parameters $x_0$. The results presented in the main text do not change when we set $x_0=0$ except for the distribution of waiting times, where we find Pareto as the best distribution if $x_0 = 0$.\\

\begin{table}[h!]
\begin{tabular}{lrrrr}
\hline 
 & Shift (Lognormal) & Shift (Pareto) & Scale (Pareto) \\
Displacements & 2.02 m & -11.41 m & 7431.83 m \\
Waiting times & 0.18 h& -0.03 h & 13.03 h\\
Discoveries & 1.9 m& -1.34 m & 2801.35 m\\
\hline
\end{tabular}
\caption{\textbf{The scale and shift parameters.} The values of the shift parameter of the Lognormal fit (first column), the shift and scale parameter of the Pareto fit of the distributions' tails (second and third columns). }
\label{TableShiftScale}
\end{table}

%
%
\subsection*{Further analysis: Selection of the best model among 68 distributions}
\begin{figure}[h!]
\centering
\includegraphics[width=\textwidth]{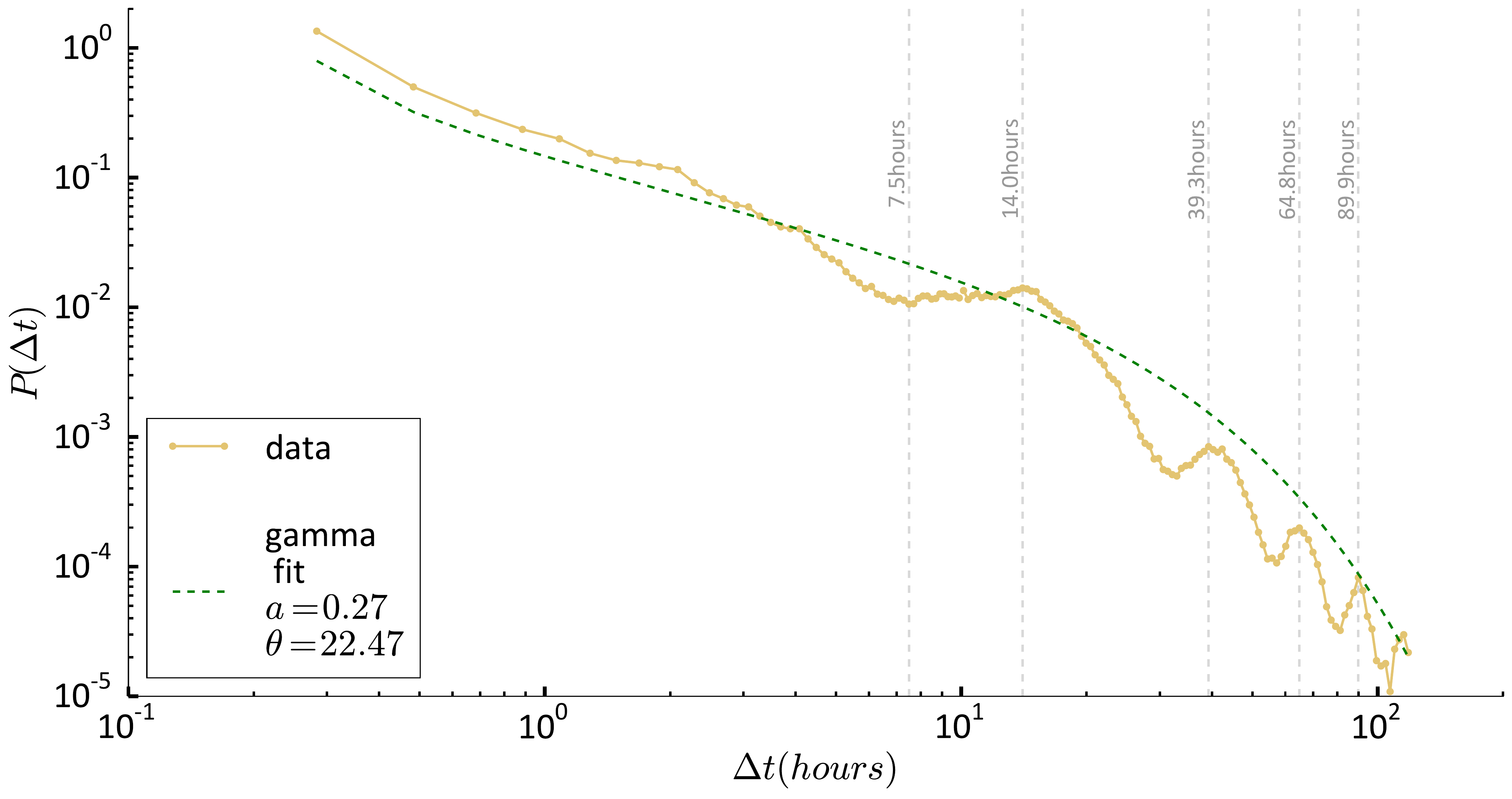}
\caption{\textbf{Distribution of waiting times: selection of the best model among 68 distributions}. Yellow dotted line: data. Green dashed line: Gamma Distribution fit with characteristic parameters $a= 0.27$ and $\theta=22.47$}
\label{FigGamma}
\end{figure}

In the case of the distribution of waiting times, the best model among 68 distributions is the gamma distribution. Results of the gamma fit are shown in figure \ref{FigGamma}.

\clearpage
\subsection*{Distributions}
The list of distributions is based on the scipy.stats Python \cite{scipystats} module which contains the implementation of over 80 probability distributions, including those reported in the literature on human mobility. We have excluded distributions with more than 3 parameters (including scale and shift), unless they were found in previous studies on human mobility. The distribution considered are the following:

\noindent \textit{Levy alpha-stable, Anglit, arcsine, Bradford, Cauchy, chi, chi-squared, cosine, double gamma, double Weibull, exponential, exponential power, fatigue-life, Fisk, folded Cauchy, folded normal, Frechet left, Frechet right, gamma, generalized extreme value, generalized Gamma, generalized half-logistic, generalized logistic, Generalized Pareto, Gilbrat, Gompertz, left-skewed Gumbel, right-skewed Gumbel, half-Cauchy, half-logistic, half-normal, hyperbolic secant, inverted gamma, inverse Gaussian, inverted Weibull, General Kolmogorov-Smirnov, Laplace, Levy, left-skewed Levy, log gamma, logistic, log-Laplace, lognormal, Lomax, Maxwell, Nakagami, normal, Pareto, Pearson type III, power-function, power log-normal , power normal, Rayleigh, R, Reciprocal inverse Gauss, Rice, semicircular, Student’s T, triangular, truncated exponential, truncated normal, Tukey-Lambda, Truncated Pareto, Uniform, Von Mises, Wald, Weibull maximum, Weibull minimum, wrapped Cauchy}\\

\end{document}